\documentclass[sigconf]{acmart}
\AtBeginDocument{%
  }

\setcopyright{acmlicensed}
\copyrightyear{2026}
\acmYear{2026}
\setcopyright{cc}
\setcctype{by}
\acmConference[FORGE '26]{2026 IEEE/ACM Third International Conference on AI Foundation Models and Software Engineering}{April 12--13, 2026}{Rio de Janeiro, Brazil}
\acmBooktitle{2026 IEEE/ACM Third International Conference on AI Foundation Models and Software Engineering (FORGE '26), April 12--13, 2026, Rio de Janeiro, Brazil}
\acmPrice{}
\acmDOI{10.1145/3793655.3793732}
\acmISBN{979-8-4007-2477-0/2026/04}

\usepackage{amsfonts}
\usepackage{algorithmic}
\usepackage{textcomp}
\usepackage{url}
\usepackage{cleveref}
\usepackage{tabularx}
\usepackage{multirow}
\usepackage[x11names,table]{xcolor}
\usepackage{colortbl}
\usepackage{makecell}
\usepackage{subcaption}

\usepackage[most]{tcolorbox}
\usetikzlibrary{shapes.geometric, positioning, fit, backgrounds, calc, shadows, trees, arrows.meta}

\newcommand{\react}{\textsc{ReAct}}
\newcommand{\pe}{\textit{Plan-and-Execute}}
\newcommand{\baseline}{\textit{Straight-Shot}}
\newcommand{\baselinetab}{\makecell{\textit{Straight-Shot}\\(Baseline)}}
\newcommand{\rfa}{Procedural}
\newcommand{\rfrca}{RCA-specific}
\newcommand{\rfg}{General}
\newcommand{\random}{Random Guessing}

\newcommand{\A}{$\mathcal{A}$}
\newcommand{\B}{$\mathcal{B}$}
\newcommand{\rqone}{$\textit{RQ}_1$}
\newcommand{\rqtwo}{$\textit{RQ}_2$}
\newcommand{\rqthree}{$\textit{RQ}_3$}
\newcommand{\rqfour}{$\textit{RQ}_4$}

\definecolor{lrgcolor}{HTML}{f0f0f0}
\newcommand{\lrg}{\cellcolor{lrgcolor}}
\newcommand{\lrgtext}[1]{\setlength{\fboxsep}{1pt}\colorbox{lrgcolor}{#1}}

\AtBeginDocument{%
  \setlength{\abovedisplayskip}{2pt plus 1pt minus 1pt}%
  \setlength{\belowdisplayskip}{2pt plus 1pt minus 1pt}%
  \setlength{\abovedisplayshortskip}{1.5pt plus 1pt minus 1pt}%
  \setlength{\belowdisplayshortskip}{1.5pt plus 1pt minus 1pt}%
}

\newcommand{\rqi}{How effectively can an LLM agent perform RCA tasks?}
\newcommand{\rqii}{How sensitive are final RCA outcomes to alert modalities (logs, metrics, traces)?}
\newcommand{\rqiii}{What reasoning failures appear in agent inference traces?}
\newcommand{\rqiv}{How does the presence of reasoning failures affect the likelihood of generating correct RCA hypotheses?}

\newcommand*{\conclusionbox}[1]{%
\noindent
#1
}

\newcommand*{\smallsection}[1]{\textbf{\emph{#1}}}
\newcounter{observatione}
\newcommand{\observation}[1]{%
 \smallsection{Observation \arabic{observatione}\stepcounter{observatione}:~#1}%
}

\begin{document}

\title{Stalled, Biased, and Confused: Uncovering Reasoning Failures in LLMs for Cloud-Based Root Cause Analysis}
%
%
%

\author{Evelien Riddell}
\email{evelien.riddell@uwaterloo.ca}
\affiliation{%
  \institution{University of Waterloo, Canada}
  \country{}
}
\author{James Riddell}
\email{james.riddell@uwaterloo.ca}
\affiliation{%
  \institution{University of Waterloo, Canada}
  \country{}
}
\author{Gengyi Sun}
\email{gengyi.sun@uwaterloo.ca}
\affiliation{%
  \institution{University of Waterloo, Canada}
  \country{}
}
\author{Michał Antkiewicz}
\email{michal.antkiewicz@uwaterloo.ca}
\affiliation{%
  \institution{University of Waterloo, Canada}
  \country{}
}
\author{Krzysztof Czarnecki}
\email{krzysztof.czarnecki@uwaterloo.ca}
\affiliation{%
  \institution{University of Waterloo, Canada}
  \country{}
}

\renewcommand{\shortauthors}{Riddell et al.}

\begin{abstract}
Root cause analysis (RCA) is essential for diagnosing failures within complex software systems to ensure system reliability. The highly distributed and interdependent nature of modern cloud-based systems often complicates RCA efforts, particularly for multi-hop fault propagation, where symptoms appear far from their true causes.
Recent advancements in Large Language Models (LLMs) present new opportunities to enhance automated RCA. 
However, their practical value for RCA depends on the fidelity of reasoning and decision-making.
Existing work relies on historical incident corpora, operates directly on high-volume telemetry beyond current LLM capacity, or embeds reasoning inside complex multi-agent pipelines---conditions that obscure whether failures arise from reasoning itself or from peripheral design choices. 

We present a focused empirical evaluation that isolates an LLM's reasoning behavior.
We design a controlled experimental framework that foregrounds the LLM by using a simplified experimental setting.
We evaluate six LLMs under two agentic workflows (\react~and \pe) and a non-agentic baseline on two real-world case studies (GAIA and OpenRCA). 
In total, we executed 48,000 simulated failure scenarios, totaling 228 days of execution time.
We measure both root-cause accuracy and the quality of intermediate reasoning traces. We produce a labeled taxonomy of 16 common RCA reasoning failures and use an LLM-as-a-Judge for annotation.
Our results clarify where current open-source LLMs succeed and fail in multi-hop RCA, quantify sensitivity to input data modalities, and identify reasoning failures that predict final correctness. 
Together, these contributions provide transparent and reproducible empirical results and a failure taxonomy to guide future work on reasoning-driven system diagnosis.
%

\end{abstract}




\maketitle

\section{Introduction} \label{sec:introduction}

Modern cloud-native, microservice-based systems promise scalability, flexibility, and resilience but pose significant challenges for reliability and observability.
In such distributed environments, a single failure can cascade across multiple components, making \textit{root cause analysis} (RCA)---the task of identifying the originating fault behind a system failure---both essential and complex. 

{\looseness=-1
RCA in cloud-based systems typically involves analyzing telemetry data (e.g., logs, metrics, traces)~\cite{Sridharan:Observability:2018, Soldani:survey:ACMCS:2022} to trace failure propagation and isolate the faulty component. However, the sheer scale, heterogeneity, and volume of telemetry data 
in production environments 
can overwhelm automated tools and human operators alike~\cite{Chen:RCACopilot:EuroSys:2024}. 
In response, recent work has explored data-driven approaches to efficiently surface informative clues that guide manual diagnosis 
by reliability engineers
or software operators
toward the true root cause~\cite{Soldani:survey:ACMCS:2022}.
\par}

The recent emergence of Large Language Models (LLMs) introduces a promising new paradigm for automated or human-in-the-loop RCA. 
Thanks to their general reasoning and tool-use capabilities, LLMs offer the potential to navigate system knowledge and extract insights from both structured (e.g., incident reports) and unstructured data (e.g., logs)~\cite{Wang:survey:arXiv:2024}. This makes them attractive candidates for autonomous RCA agents capable of integrating system information with observed system behavior.
However, important methodological and evaluative gaps remain that prevent us from understanding how adept LLMs \textit{actually} are for RCA.

\textbf{First, methodological complexity in framework designs obscures core reasoning.}
Many approaches embed LLMs within complex agentic workflows or multi-agent frameworks, where RCA outputs reflect the combined contributions of multiple components (e.g., coordinator, analyzer, code-executor)~\cite{Wang:RCAGent:CIKM:2024,Zhang:mABC:EMNLP:2024,Xu:OpenRCA:ICLR:2025}.
This entanglement makes it difficult to assess individual agent contributions, particularly when discerning between the primary reasoning thread and auxiliary tasks or inter-agent chatter.

\textbf{Second, no prior work systematically evaluates reasoning quality for the RCA task.}
Existing studies primarily report final-output accuracy or efficiency, and are sometimes supplemented by coarse-grained human judgments~\cite{Zhang:mABC:EMNLP:2024, Goel:LLM-InContext:FSE:2024, Roy:LLM-React:FSE:2024}.
This lack of systematic, reproducible, and automated evaluation of reasoning quality constrains researchers from conducting large-scale evaluations and comparisons across studies.

These limitations point to a broader gap: despite the increasing adoption of LLMs for RCA, we lack a clear, empirical understanding of their isolated reasoning capabilities, the factors influencing their performance, and the reasoning failures they exhibit.
To address this gap, we conduct a systematic empirical evaluation in a controlled RCA setting that isolates the LLM’s reasoning behavior from confounding factors 
(e.g., inter-agent dependencies, system-specific heuristics, multiple sequential/overlapping tasks).
Unlike prior multi-agent or heavily engineered frameworks, we use simple agent architectures, deterministic tools, alert-level inputs, and an explicit typed knowledge graph to foreground reasoning behavior.
This simplified design enables targeted analysis of both final RCA outputs and intermediate reasoning traces.

We focus our study with the following research questions:
\\
\textbf{\rqone}: \rqi \\
\textbf{\rqtwo}: \rqii \\
\textbf{\rqthree}: \rqiii \\
\textbf{\rqfour}: \rqiv

{\looseness=-1
Our \textbf{main contributions} are:
(i) an evaluation framework that isolates LLM-based RCA reasoning performance from confounding factors;
(ii) a human-rated LLM-as-a-Judge evaluator capable of assessing RCA reasoning quality automatically and reproducibly;
(iii) an extensive empirical study (48,000 samples; 228 days of execution time) on the capabilities of LLMs for RCA under several agentic and one non-agentic workflows;
(iv) and a complete replication package~\cite{appendix} with our implementation, evaluation setup, and results.
This study is the first empirical investigation into the isolated ability of LLM-based RCA agents and their reasoning failures.
For additional information and discussion of these findings, see~\cite{thesis}.
\par}
\section{Related Work \& Problem Definition}
\begin{figure*}[t]
    \centering
    \includegraphics[width=0.99\textwidth]{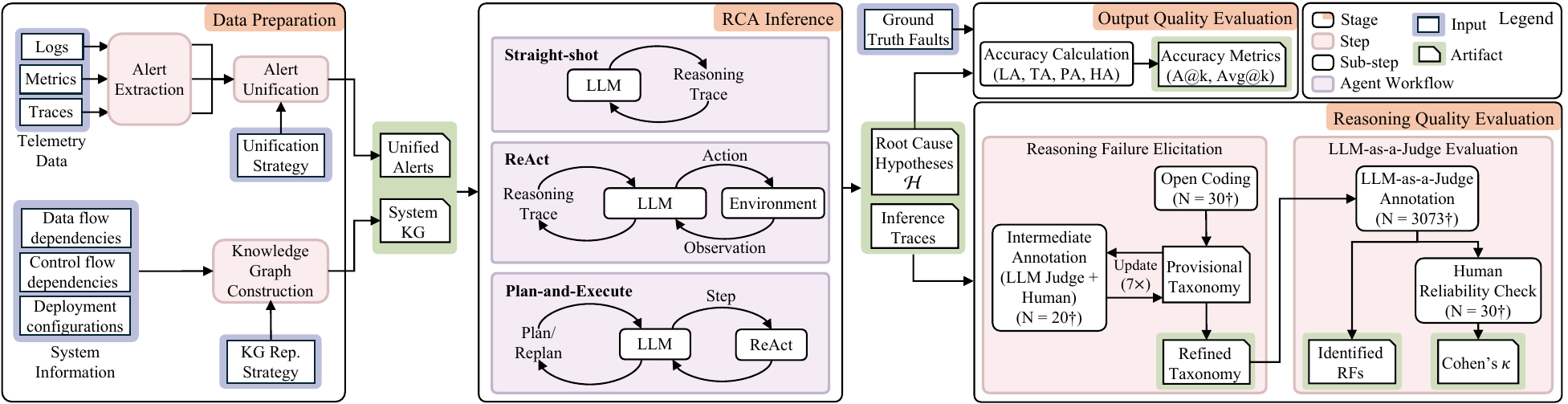} 
    \vspace{4pt}
    \caption{Overview of the study method.}
    \label{fig:method-overview}
\end{figure*}

In this section, we present related work 
on LLM-based RCA (\Cref{subsec:LLMRCA}) and RCA evaluation (\Cref{subsec:RCAE}) 
and define our problem scope (\Cref{sec:problem-def}).

\vspace{-6pt}
\subsection{LLM-based RCA} \label{subsec:LLMRCA}
{\looseness=-1
The promise of LLMs has led to a surge of interest in applying them to RCA
\cite{Ahmed:LLM-Finetune:ICSE:2023, Wang:RCAGent:CIKM:2024, Zhang:LLM-InContext:FSE:2024, Roy:LLM-React:FSE:2024, Chen:RCACopilot:EuroSys:2024, Goel:LLM-InContext:FSE:2024, Zhang:mABC:EMNLP:2024, Han:LasRCA:ASE:2024, Pei:Flow-of-Action:WWW:2025, Xu:OpenRCA:ICLR:2025, Qiu:RAG-LLM-Hardware:COINS:2025}.
Early approaches have explored fine-tuning~\cite{Ahmed:LLM-Finetune:ICSE:2023} or in-context learning~\cite{Zhang:LLM-InContext:FSE:2024, Goel:LLM-InContext:FSE:2024, Roy:LLM-React:FSE:2024, Chen:RCACopilot:EuroSys:2024} for domain adaptation, typically using labeled historical incident reports. 
While these methods effectively leverage LLMs' pattern-matching and information-aggregation abilities, they provide the LLM with a limited (and often summarized) view of the system, potentially failing to capture the complexities of entity interactions or provide useful diagnostic signals (e.g., multi-modal telemetry data). Moreover, limiting LLM input to static or historical data can hinder its ability to adapt to the system's current state and evolving interactions.
\par}

{\looseness=-1
More recent work integrates LLMs as agents into retrieval- or tool-augmented frameworks~\cite{Roy:LLM-React:FSE:2024, Qiu:RAG-LLM-Hardware:COINS:2025, Zhang:TAMO:TSC:2025} or multi-agent systems~\cite{Wang:RCAGent:CIKM:2024, Zhang:mABC:EMNLP:2024, Han:LasRCA:ASE:2024, Pei:Flow-of-Action:WWW:2025, Xu:OpenRCA:ICLR:2025}. 
These workflows generally enable the LLM to retrieve additional system or diagnostic information, such as service dependencies~\cite{Zhang:mABC:EMNLP:2024}, more detailed incident information~\cite{Roy:LLM-React:FSE:2024, Wang:RCAGent:CIKM:2024}, similar historical incidents~\cite{Roy:LLM-React:FSE:2024}, relevant code snippets and execution paths~\cite{Li:COCA:ICSE:2025}, or other LLM expert agents (e.g., as analytical tools like code~\cite{Wang:RCAGent:CIKM:2024} or fault probability analyzers~\cite{Zhang:mABC:EMNLP:2024}, or python executors~\cite{Xu:OpenRCA:ICLR:2025}).
Although these approaches show promise, their architectural complexity introduces confounding factors that make it difficult to isolate and assess LLMs’ actual ability to reason about RCA tasks.
\par}

\vspace{-6pt}
\subsection{RCA Evaluation}  \label{subsec:RCAE}
{\looseness=-1
\textbf{The capabilities of LLMs are often obscured in existing LLM-based RCA approaches},
making it difficult to isolate and evaluate how and why an approach arrives at a particular diagnosis.
Existing approaches frequently combine several sequential tasks (e.g., anomaly detection and fault analysis) into a single pipeline~\cite{Han:LasRCA:ASE:2024, Xu:OpenRCA:ICLR:2025} or distribute sub-tasks across specialized agents~\cite{Han:LasRCA:ASE:2024, Wang:RCAGent:CIKM:2024, Zhang:mABC:EMNLP:2024, Pei:Flow-of-Action:WWW:2025, Xu:OpenRCA:ICLR:2025}.
For example, OpenRCA~\cite{Xu:OpenRCA:ICLR:2025} frames RCA as a sequence of subtasks (i.e., anomaly detection, failure identification, and fault analysis) and relies on open-ended code execution and telemetry analysis.
This both introduces multiple potential failure sources and conflates the core RCA ability with unrelated skills such as anomaly detection, code generation, and error handling.
Other systems incorporate static protocols or incident-specific handlers
\cite{Chen:RCACopilot:EuroSys:2024}.
While such setups show promise, these design choices entangle reasoning with opaque task boundaries, inter-agent dependencies, and system-specific heuristics, making it difficult to attribute performance bottlenecks or errors to the LLM. This challenge is especially pronounced in multi-hop scenarios where intermediate reasoning steps matter. To assess LLMs as effective RCA agents, we require evaluation settings where reasoning behavior can be meaningfully surfaced and analyzed in isolation.
\par}

{\looseness=-1
\textbf{Existing evaluation metrics overlook reasoning quality and propagation correctness.}
Current evaluations of RCA agents are often restricted to final output or aggregate human assessments---e.g., perceived ``usefulness''~\cite{Zhang:mABC:EMNLP:2024}, overall ``correctness''~\cite{Goel:LLM-InContext:FSE:2024}, or post-hoc ``reasoning error'' counts~\cite{Roy:LLM-React:FSE:2024}, offering little transparency and interpretability.
Many rely on lexical or semantic similarity between predicted and reference root-cause descriptions~\cite{Ahmed:LLM-Finetune:ICSE:2023, Wang:RCAGent:CIKM:2024, Zhang:LLM-InContext:FSE:2024, Roy:LLM-React:FSE:2024, Goel:LLM-InContext:FSE:2024}, which can mask partially incorrect diagnoses and offer little interpretability. 
These metrics are particularly inadequate in multi-hop RCA settings, where correctness depends not only on the final root cause but also on the inferred fault propagation path. 
While mABC~\cite{Zhang:mABC:EMNLP:2024} incorporates propagation path evaluation, its multi-agent setup complicates attribution, making it difficult to disentangle the primary reasoning decisions from those of downstream agents.
To understand and improve LLM performance as RCA agents, we need evaluation protocols that expose not only what was predicted, but how and why it was inferred.
\par}

\vspace{-6pt}
\subsection{Problem Definition}\label{sec:problem-def}
To address these limitations, we formulate RCA as a reasoning-focused task centered on three objectives: (1) root cause localization, (2) fault type classification, and (3) propagation path identification. 
Given a set of multi-modal monitoring alerts triggered by a single fault, the LLM agent is tasked with jointly performing these objectives and producing a ranked list of $k$ fault hypotheses.
Each hypothesis consists of a candidate root-cause entity, an associated fault type, an inferred fault propagation path through the system KG, and a natural-language justification.

\textbf{A simplified, structured environment enables more interpretable and better reasoning evaluation.}
By isolating the LLM, our framework serves as a testbed for evaluating whether models can meaningfully plan, adapt, and infer during fault diagnosis.
Importantly, we argue that a simplified RCA setting does not trivialize the RCA task: multi-hop failure propagation across distributed services still presents significant reasoning challenges. 
Rather, the controlled setup reduces confounding factors (e.g., noisy telemetry, multi-agent contribution, lack of clear task separation), thereby foregrounding reasoning behavior and offering a clearer window into agent behavior.
Coupled with an explicit system KG, this setup enables agents to reason over structured and inspectable entity relationships, supporting transparency and verifiability. 
We assume pre-identified anomaly events and focus exclusively on the post-detection stage of RCA, shifting attention away from data wrangling or auxiliary analysis tasks to a systematic evaluation of both output accuracy, propagation inference, and intermediate reasoning quality.
\section{Study Methodology} \label{sec:methodology}

\Cref{fig:method-overview} summarizes the end-to-end overview of our study methodology.
We first extract modality-specific alerts from raw telemetry data (logs, metrics, and traces)---which are canonicalized into a unified alert format---and a KG that encodes the entities, control- and data-flow dependencies, and deployment configurations of the software system (\Cref{sec:data-preparation}).
These form the input into the RCA inference setup (\Cref{sec:rca-inference}), where an LLM is run under three distinct workflows (\baseline, \react, \pe), producing final root-cause hypotheses (location, fault type, propagation path) and step-level inference traces.
Finally, we evaluate both the final outputs and the inference traces (\Cref{sec:evaluation}).
We expand on each stage in the following sections.

\vspace{-6pt}
\subsection{Data Preparation} \label{sec:data-preparation}

\begin{figure}[!t]
    \centering
    \includegraphics[width=\linewidth]{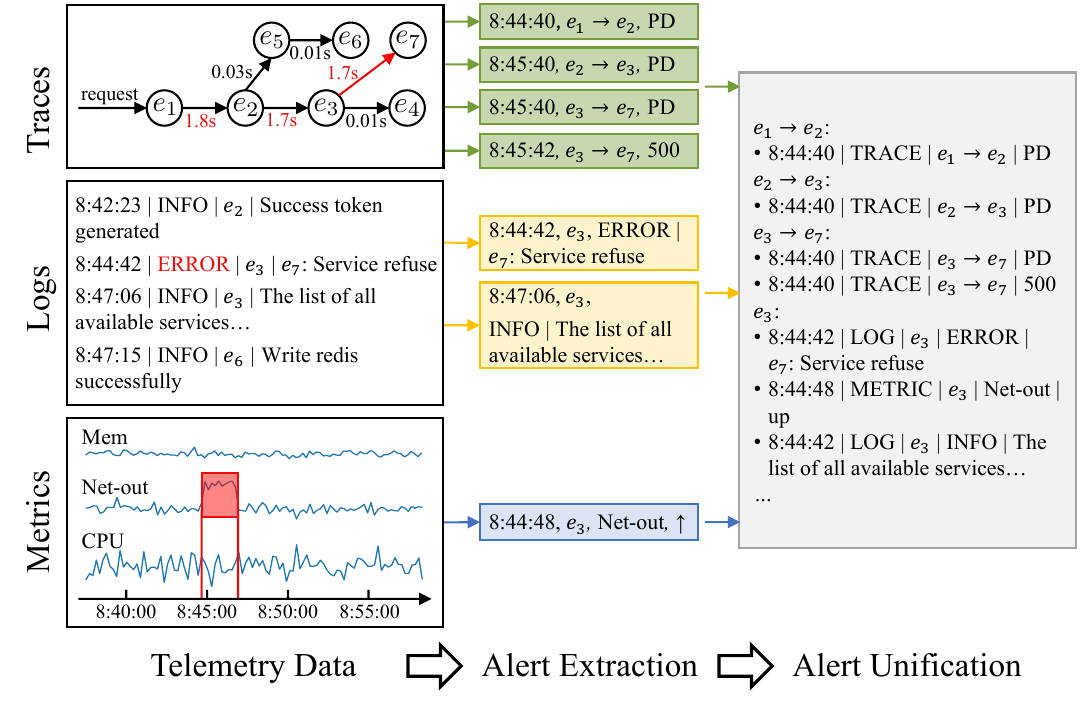} 
    \caption{The alert extraction and unification process using traces, logs, and metrics.}
    \label{fig:alert-extraction}
\end{figure}

Data preparation consists of alert compilation and KG construction. The former is decomposed into two sequential steps: 1) alert extraction and 2) alert unification.
~\Cref{fig:alert-extraction} shows an example of alert compilation for a single fault.

\vspace{-6pt}
\subsubsection{Alert Extraction} \label{sec:alert-extraction}
We perform multi-modal alert extraction as a data pre-processing step. 
This step is designed to emulate realistic production settings, where monitor alerts or event triggers serve as the primary entry point to failure triage and diagnosis~\cite{Goel:LLM-InContext:FSE:2024}.
Since most open-source RCA monitoring datasets consist of large quantities of raw telemetry without clear natural language alert labels, we extract and process the alert events suitable for LLM use.
%
%
Specifically, we adopt a \textit{feature fusion} approach~\cite{Zhang:survey:TSEM:2025}, where we process multi-modal monitoring data and extract a unified representation of the fault---serving as input to the LLM agent. 
Our approach follows prior work on fusion-based approaches~\cite{Zhang:DiagFusion:TSC:2023, Guangba:Nezha:FSE:2023, Xie:TVDiag:arXiv:2025}, applying anomaly detection techniques independently to each modality (metrics, traces, and logs) to generate alerts.

\noindent
\textbf{Log Alerts.} We use a
log parsing approach Drain~\cite{He:Drain:ICWS:2017} to extract static log templates and dynamic log parameters from raw log messages, similar to~\cite{Guangba:Nezha:FSE:2023, Zhang:DiagFusion:TSC:2023, Xie:TVDiag:arXiv:2025}.
To ensure that we select valuable logs for failure diagnosis and filter out unnecessary noise, we use a two-part log alert sampling technique. 
First, we preserve occurrences of ERROR-level and low-frequency log templates as alerts, following the approach in~\cite{Xie:TVDiag:arXiv:2025}.
Second, to ensure that the alerts fit in the LLM context window, we collapse high-volume log template instances into a ``representative'' log message associated with a given system entity---specifically, the first occurrence. 

\noindent
\textbf{Trace Alerts.} 
We use an Isolation Forest (IForest)~\cite{Liu:IsolationForest:2008} to detect anomalies in traces between services and service instances. IForest is an unsupervised anomaly detection method that isolates anomalies by recursively partitioning data using random splits; anomalous points are more susceptible to isolation and thus require fewer partitions.
Specifically, we apply IForest to the response time and status codes of each invocation pair to identify deviations from normal behavior. 
%
High response time indicates potential performance degradation (PD), while abnormal status codes suggest errors (ERROR). 

\noindent
\textbf{Metric Alerts.}
We opt for the 3-sigma rule~\cite{Pukelsheim:3sigma:1994} to detect anomalous metrics. For a given metric, we collect the numerical fluctuations over a time period and compute the mean $\mu$ and standard deviation $\sigma$. When the value of a metric exceeds the upper bound of 3-sigma ($\mu + 3 \sigma$), it is deemed an alert with an \textit{up} anomaly direction. Similarly, the anomaly direction is \textit{down} for an alert if the metric value falls below the lower bound ($\mu - 3 \sigma$). 

\vspace{-6pt}
\subsubsection{Alert Unification} \label{sec:alert-unification}
To obtain a consistent input representation for the LLM, we group alerts according to a \textit{unification strategy}.
For input robustness, we consider two strategies: (1) \textit{time-based unification}, which merges alerts into a single chronological sequence;
and (2) \textit{element-based unification} (shown in~\Cref{fig:alert-extraction}), which aggregates alerts by their reporting element.

\vspace{-6pt}
\subsubsection{System Knowledge Graph (KG) Construction} \label{sec:knowledge-graphs}


{\looseness=-1
Most existing research relies on incident metadata (e.g., title and summary) for fine-tuning LLMs~\cite{Ahmed:LLM-Finetune:ICSE:2023} or directly querying LLMs with in-context examples~\cite{Zhang:LLM-InContext:FSE:2024, Goel:LLM-InContext:FSE:2024, Roy:LLM-React:FSE:2024, Chen:RCACopilot:EuroSys:2024}.
However, failures stemming from faults in dependent components require additional information for holistic reasoning. Both~\cite{Goel:LLM-InContext:FSE:2024} and~\cite{Zhang:mABC:EMNLP:2024} show that upstream dependency information can assist LLMs in better reasoning and improve the quality of recommendations. 
To this end, we model a software system as an explicit, typed KG that captures its structural and operational aspects. KGs' semantic richness is well-suited for integration with LLMs for relational reasoning and fault-propagation analysis~\cite{Wu:LLM-GR:arXiv:2024}.
\par}

{\looseness=-1
The graph nodes represent system entities consisting of both software (e.g., services, data stores) and hardware components (e.g., hosts), while relationships capture dependencies and interactions commonly observed in cloud-native architectures, such as control and data flow dependencies, instantiations of software components 
(e.g., \textit{instance-of}), and deployment relations (e.g., \textit{hosted-on}). 
Unlike raw distributed-tracing graphs or simple dependency trees, a typed KG explicitly encodes heterogeneous entity types and relationship types (e.g., \textit{service}/\textit{host}; \textit{control-flow}/\textit{hosted-on}).
\par}

{\looseness=-1
To build the KG, we manually parse architectural documentation, deployment manifests, and configuration files, and normalize the results into a typed entity-and-relationship schema.
We represent the KG according to a \textit{KG representation strategy}. Similar to the alert unification strategy, the purpose of this is to increase the robustness of the input prompts to model-specific biases. We use two strategies: (1) \textit{list representation}, where nodes and edges are represented as bullet-form lists, and (2) \textit{JSON representation}, where nodes and edges are provided as a JSON object.
\par}

\vspace{-6pt}
\subsection{Root Cause Analysis Inference} \label{sec:rca-inference}

We evaluate LLM-based agents responsible for diagnosing multi-hop failures under three paradigms: reactive reasoning with tool-based feedback (\react), plan-then-execute reasoning (\pe), and a non-agentic baseline. 
These paradigms differ primarily in how the agent organizes decision-making and interacts with an external environment.
In this section, we further describe the (1) agent workflows, (2) available tools, and (3) input prompts.

\vspace{-6pt}
\subsubsection{Workflows} \label{sec:workflows}
All workflows involve a single LLM-based agent operating autonomously.

\noindent
\textbf{\react.} The agent follows a classical \react~(Reasoning + Acting)~\cite{Yao:ReAct:ICLR:2023} pattern, interleaving tool-based observations with incremental reasoning. 
Concretely, the agent implements a \textit{thought}-\textit{action}-\textit{observation} loop: it reasons, issues a tool call (action), receives an observation, and uses that observation to inform the next reasoning step.
This fine-grained, reactive workflow closely mirrors real-world RCA, i.e., sequential troubleshooting decisions and knowledge-intensive analysis of newly acquired information. The interleaved structure allows the agent to quickly adapt to new information without committing to a fixed long-horizon plan upfront. 

\noindent
\textbf{\pe.} 
The \pe~workflow follows the Plan-and-Solve~\cite{Wang:PlanSolve:ACL:2023} prompting paradigm, where planning and execution are explicitly separated to encourage more deliberate exploration, clearer intermediate outputs, and reduced redundant reasoning loops.
The agent first generates a high-level investigative plan from the initial scenario context, then executes each step sequentially using available tools.
Based on the results of each step, the plan is adjusted accordingly by the LLM.
Effective up-front planning for RCA remains a challenge for LLMs~\cite{Pei:Flow-of-Action:WWW:2025}, motivating our evaluation of this separation between planning and executing.

\noindent
\textbf{\baseline~(Non-agentic Baseline).} The non-agentic baseline implements a straight-shot workflow: the LLM receives all alerts and system knowledge upfront and returns a diagnosis in a single pass, without intermediate tool use. This straight-shot setup mirrors direct-answering workflows where the model must reason in one pass, optionally exposing chain-of-thought~\cite{Wei:CoT:NeurIPS:2022} or ``think" traces.

\vspace{-6pt}
\subsubsection{Tools} \label{sec:tools}

We design the tools provided to the LLM agent to be semantically minimalist (i.e., each tool has narrow, clearly defined semantics) and deterministic. We refrain from allowing direct access to the KG or code-writing capabilities for exploration or analysis, inspired by observations from~\cite{Wang:RCAGent:CIKM:2024, Xu:OpenRCA:ICLR:2025}. 
Although this reduces the action space, it ensures a more reliable and interpretable evaluation of reasoning behavior. Specifically, it confines agent decisions to a well-defined, reproducible set of tool calls, making it possible to attribute performance outcomes to reasoning quality rather than tool ambiguity, stochasticity, or code execution artifacts.
To this end, the tools serve to expose entity relationships and the system topology. They are organized into two categories, \textit{Data Characteristics} and \textit{Graph Traversal}.
For every tool call, we require the LLM to provide a justification for its invocation. We do this by adding an additional required parameter ``reasoning" to each tool definition.


\emph{Data Characteristics Tools} allow the agent to validate the presence of specific entities, inspect their attributes and alerts, list all instances of a given type, and examine attributes of a given relationship. These tools enable structured inspection of relevant entities and alert data. \emph{Graph Traversal Tools} allow the agent to explore the structural topology of the system. The agent can retrieve the $r$-hop neighborhood of a given entity, or query for all simple paths between two entities. These operations support inference over causal chains and fault propagation paths.

\vspace{-6pt}
\subsubsection{Input Prompts} \label{sec:input-prompts}

Each agent workflow is guided by an initial prompt that defines the RCA objective (generating a ranked list of \textit{k} fault hypotheses), KG schema (entity and relationship types), and available tools.
It also contains the unified alert representation (\Cref{sec:data-preparation}) for a particular fault scenario. 
While the core components of the prompts---task objectives, schema, output requirements, alert data---remain consistent across workflows, each prompt is tailored to the respective decision-making strategy (e.g., incremental reasoning in \react, plan specification in \pe). Additionally, we provide the full KG in text format for the \baseline~approach.

\vspace{-6pt}
\subsection{Evaluation} \label{sec:evaluation}
To assess LLM capabilities, we assess both the final root-cause hypotheses and the inference traces produced during the RCA inference stage.
We define an \textit{inference trace} to consist of all reasoning traces (i.e., intermediate LLM outputs, content of think tags), actions (i.e., tool calls and their justification), observations (i.e., tool outputs), and plan steps (for \pe~case). This forms a linear history of thought (and action) that led to the final root-cause hypotheses produced by the LLM agent.

\vspace{-6pt}
\subsubsection{Output Quality} \label{sec:metrics}
To answer \rqone~and \rqtwo,
we assess the alignment of each root-cause hypothesis with the ground-truth to evaluate the ability of the LLM to \textit{correctly} identify the root cause.
As such, we measure the root-cause Location Accuracy (LA), Type Accuracy (TA), and Hypothesis Accuracy (HA), where HA requires both fault location and type to match the ground truth.
Additionally, we measure Propagation Path Validity (PA), which checks whether the predicted propagation path is a valid walk in the KG. For each accuracy measure, we employ the top-$k$ accuracy (A@$k$), which checks the presence of the ground truth in the top-$k$ hypotheses. Formally, let $\mathcal{S}$ be the set of system faults, $s \in \mathcal{S}$ denote one fault case, $V_s$ denote the ground-truth value of $s$, and $H_s[k] = \{h_s^{(i)}\}_{i=1}^k$ denote the top-$k$ predicted hypotheses for $s$. Then,  
\begin{equation}
    \text{A@}k = \frac{1}{|\mathcal{S}|} \sum_{s \in \mathcal{S}} 
    \begin{cases}
        1, & \text{if } V_s \in \mathcal{H}_s[k] \\
        0, & \text{otherwise}
    \end{cases}
\end{equation}
We report results for $k=1, 3$.
We additionally measure the Average Accuracy@$K$ (Avg@$K$) for $k=3$, defined as
\begin{equation}
    \text{Avg@}K = \frac{1}{K} \sum_{1 \leq k \leq K} \text{A@}k.
\end{equation}


\vspace{-6pt}
\subsubsection{Reasoning Quality} \label{sec:reasoning-quality}
{\looseness=-1
Evaluating the quality of an LLM’s reasoning requires methods that can capture both the depth and nuance of its decision process. 
In our setting, RCA inference traces are long, unstructured, and interleaved with tool interactions, making them unsuitable for structured parsing approaches (e.g., first-order logic, causal graphs)
and conventional reasoning metrics (e.g., ROSCOE~\cite{Golovnenva:Roscoe:ICLR:2023}, ReCEval~\cite{Prasad:ReCEval:EMNLP:2023}, BERTScore~\cite{Zhang:BERTScore:ICLR:2020}).
Instead, we adopt a rationale-based evaluation scheme~\cite{Mondorf:ReasoningSurvey:COLM:2024} that combines qualitative analysis with structured LLM-based judging (i.e., \textit{LLM-as-a-Judge}~\cite{Zheng:LLMJudge:NeurIPS:2023}). 
This hybrid approach offers both interpretability and scalability: qualitative inspection grounds the evaluation in observed reasoning patterns, while LLM-based judging enables greater systematic coverage.
\Cref{fig:method-overview} summarizes the procedure. 
Nodes labeled with N\textsuperscript{†} denote independent random samples of size N.
\par}

\noindent
\textbf{Reasoning Failure Elicitation.} \label{sec:reasoning-failure-elicitation}
\begin{table*}[t]
\centering
\footnotesize
\renewcommand{\arraystretch}{0.95}
\caption{Reasoning Failure Taxonomy}
\label{tab:reasoning-failures-taxonomy}
\begin{tabular}{@{}lllll@{}}
\toprule
\textbf{ID} & \textbf{Name} & \textbf{Description} & \textbf{Scope} & \textbf{Category} \\
\midrule
RF-01 & Fabricated evidence & Asserts existence of alerts, metrics, logs, or traces not found in the provided evidence. & Per-hypothesis & \rfg \\
RF-02 & Metric interpretation error & Misreads metric semantics (e.g., inverts directionality or confuses counters and gauges). & Per-hypothesis & \rfrca \\
RF-03 & Confused provenance & Attributes causation to the component observing a symptom rather than its true source. & Per-hypothesis & \rfrca \\
RF-04 & Temporal misordering & Infers causal direction that violates chronological order of observed events. & Per-hypothesis & \rfg \\
RF-05 & Spurious causal attribution & Claims causal relationships unsupported by alerts or knowledge graph structure. & Per-hypothesis & \rfg \\
RF-06 & Unjustified instance specificity & Asserts instance-level fault without discriminating instance-specific evidence. & Per-hypothesis & \rfrca \\
RF-07 & Arbitrary evidence selection & Chooses evidence subsets inconsistent with systematic triage heuristics. & Per-hypothesis & \rfrca \\
RF-08 & Evidential insufficiency & Relies on weak or non-specific evidence insufficient to support the diagnostic claim. & Per-hypothesis & \rfg \\
RF-09 & Failure to update belief & Does not revise or retract claims contradicted by later evidence or tool outputs. & Full trace & \rfa \\
RF-10 & Simulation or role confusion & Treats simulated or assumed tool outputs as factual evidence. & Full trace & \rfa\\
RF-11 & Excessive speculation & Engages in prolonged speculative or circular reasoning that obstructs analysis. & Full trace & \rfa \\
RF-12 & Repetition or failure to resume & Repeats planning or reasoning across turns without substantive progress. & Full trace & \rfa \\
RF-13 & Anchoring bias & Fixates prematurely on one hypothesis and neglects exploration of alternatives. & Cross-cutting & \rfg \\
RF-14 & Invalid inference pattern & Applies formal or informal logical fallacies in diagnostic reasoning. & Cross-cutting & \rfg\\
RF-15 & Internal contradiction & Produces mutually inconsistent statements within the inference history. & Cross-cutting & \rfg \\
RF-16 & Arithmetic or aggregation error & Performs numeric miscalculations or aggregations affecting interpretation. & Cross-cutting & \rfa\\
\bottomrule
\end{tabular}
\end{table*}
 
{\looseness=-1
We developed a taxonomy of \textit{reasoning failures} (RFs) through a structured, mixed-methods elicitation procedure. 
Starting from a random sample of 30 RCA outputs (drawn from diverse agent and input configurations to maximize behavioral coverage), we performed \textit{Open Coding}~\cite{Corbin:OpenCoding:2025} to identify recurrent reasoning issues (e.g., related to evidence handling, temporal inference, provenance, causal attribution, tool misuse).
We subsequently drafted definitions and exemplars for each provisional code. These initial codes formed the \textit{Provisional Taxonomy} used in subsequent rounds.
Taxonomy refinement proceeded through seven iterative rounds.
In each round, we sampled a new random batch of 20 RCA outputs.
Each sample underwent \textit{Intermediate Annotation} in parallel by an LLM-based judge and a human annotator according to the working taxonomy.
In each round, we collected novel failure cases, updated failure definitions and examples, and refined the set of evaluation steps provided to the judge. This paired-annotation approach (LLM judge + human) was used to align interpretations and, in turn, update the taxonomy and input prompt. Only minor revisions were required in the last round, thereby forming the \textit{Refined Taxonomy} (\Cref{tab:reasoning-failures-taxonomy}), and we used this taxonomy, evaluation steps, and prompt for the final \textit{LLM-as-a-Judge Annotation}.
\par}

{\looseness=-1
We organize the failures into three interpretable buckets. \textit{Procedural} failures arise in the agent's decision-to-act loop (planning, tool invocation/interpretation, multi-turn control), causing procedural breakdowns in how the LLM functions as a reasoning agent. \textit{RCA-specific} failures are domain-related errors tied to diagnostic knowledge (metric semantics, provenance, instance discrimination, triage heuristics). \textit{General} reasoning failures are defects that undermine the credibility of hypotheses or whole traces (unsupported causal links, temporal ordering errors, anchoring, logical fallacies, internal contradictions).
\par}

\noindent
\textbf{LLM-as-a-Judge Evaluation.} \label{sec:llm-as-a-judge-evaluation}
{\looseness=-1
LLM-as-a-Judge merges the scalability of automatic methods with the detailed, context-sensitive reasoning found in expert judgments~\cite{Gu:LLMJudgeSurvey:arXiv:2025}.
We employ GPT-5~\cite{gpt5}, a state-of-the-art reasoning model, as an independent evaluator to annotate reasoning failure across RCA traces.
To mitigate self-enhancement bias---a phenomenon that LLM evaluators may prefer responses generated by themselves~\cite{Zheng:LLMJudge:NeurIPS:2023, Ye:LLMJudgeBias:ICLR:2025}---the judge model differs from those under evaluation.
Following best practices to ensure reliable and accurate evaluations, it is guided by prompts that incorporate: (1) the reasoning failure taxonomy with definitions and exemplars, (2) decomposed evaluation steps, and (3) a fixed output schema requiring categorical judgments and textual explanations with evidence from the RCA inference trace. We apply this judge to a stratified subset of 3,073 (of 19,200; 16\%) RCA inference traces ($\sim$100 per dataset-model-workflow combination) to quantify failure prevalence.
\par}

{\looseness=-1
To provide a \textit{Reliability Check} of the judge outputs, two authors independently reviewed a random sample of 30 judge-annotated traces.
Each author applied the taxonomy and we report Cohen's $\kappa=0.92$ (95\% CI) as a reliability metric. These human annotations were used only to assess judge reliability and were not used to alter final judge labels or for downstream analysis.
\par}


\vspace{-6pt}
\subsection{Experimental Setup} \label{sec:experimental-setup}

\subsubsection{Experimental Data} \label{sec:data}
\begin{table}[t]
\centering
\footnotesize
\renewcommand{\arraystretch}{0.9}
\caption{Experimental Data Details}
\label{tab:dataset-details}
\renewcommand{\arraystretch}{1.2}
\setlength{\tabcolsep}{4pt}
\begin{tabular}{@{}cccccc@{}}
\toprule
\textbf{Dataset} & \textbf{System} & \textbf{Samples} & \textbf{Entities} & \textbf{\makecell{Fault\\Categories}} & \textbf{ \makecell{Fault\\Granularity}}
\\ \midrule
    $\mathcal{A}$ & MicroSS & 152 & 24 & 5 & Service instance \\ 
    $\mathcal{B}$ & \makecell{Online\\Boutique} & 148 & 60 & 15 & \makecell{Host, service,\\service instance}\\ 
\bottomrule
\end{tabular}
\end{table}
{\looseness=-1
We conducted our experiments on two open-source microservice datasets (\A, \B) with multi-modal monitoring data (i.e., logs, metrics, and traces), listed in~\Cref{tab:dataset-details}.
Dataset \A~is derived from the Generic AIOps Atlas (GAIA) dataset~\cite{GAIA:dataset} which is collected on the MicroSS microservice application.
Dataset \B~is derived from the \textit{Market} dataset of OpenRCA~\cite{Xu:OpenRCA:ICLR:2025} which is collected from an open-source cloud-native application Online Boutique~\cite{OnlineBoutique:benchmark}.
%
%
These systems have been widely used in many previous RCA studies 
\cite{Sui:LogKG:TSC:2023, Guangba:Nezha:FSE:2023, Lee:Eadro:ICSE:2023, Zhang:DiagFusion:TSC:2023,
    Zheng:MULAN:WWW:2024, Han:LasRCA:ASE:2024, Han:HolisticRCA:TSC:2024, Wang:MRCA:ASE:2024,
    Xie:TVDiag:arXiv:2025, Xu:OpenRCA:ICLR:2025, Pei:Flow-of-Action:WWW:2025} 
and have become important experimental platforms for researchers to study microservice architectures and performance.
\par}

\vspace{-6pt}
\subsubsection{Large Language Models} \label{sec:llms}
{\looseness=-1
To support the RCA task requiring processing long contexts and action formulation,
we selected a set of open-source instruction-tuned models with 
at least 128K token context windows and support for tool calling and structured output from a variety of model families and parameters, including
Llama 3.2 (3B)~\cite{llama32},
Qwen 3 (4B and 32B)~\cite{qwen3},
Llama 3.3 Instruct (70B)~\cite{llama33}, and
Command R+ (104B)~\cite{command-r-plus}.
We additionally select a distilled DeepSeek-R1 (70B)~\cite{deepseekr1} reasoning model.
However, we evaluate Deepseek-R1 only on the \baseline~workflow as it does not support tool calling.
The model checkpoints are included in the supplementary material.
We restrict our selection to open-source models as general LLM methods for RCA and other cloud AIOPs tasks should be internally hosted for privacy and security concerns~\cite{Wang:RCAGent:CIKM:2024}. Therefore, we perform our evaluation of models generally suitable for in-house execution.
\par}

\vspace{-6pt}
\subsubsection{Implementation and Settings} \label{sec:implementation}
{\looseness=-1
We implement all agent workflows using LangGraph
\cite{langgraph}
with a graph recursion limit of 50. 
%
%
All experiments were run locally on a Linux machine with Python 3.12, across ten 48 GB NVIDIA RTX A6000 GPUs and eight 32 GB NVIDIA Tesla V100 GPUs. We executed 48,000 fault scenarios with a total execution time of 228 days.
\par}
\section{Results} \label{sec:results}

\subsection{\rqone: \rqi} \label{sec:results-rq1}
\begin{table*}
\centering
\footnotesize
\renewcommand{\arraystretch}{0.9}
\caption{Accuracy Results}
\label{tab:accuracy-results}
\setlength{\tabcolsep}{2.5pt}
\begin{tabular}{ccc|cccccc|ccc|ccc}
\toprule
Dataset & Workflow & Model & LA@1 & LA@3 & LA-Avg@3 & TA@1 & TA@3 & TA-Avg@3 & PA@1 & PA@3 & PA-Avg@3 & HA@1 & HA@3 & HA-Avg@3 \\
\midrule
\multirow[c]{17}{*}{$\mathcal{A}$} & \multirow[c]{6}{*}{\baselinetab} & Llama 3.2 & 0.13 & 0.31 & 0.23 & 0.24 & 0.64 & 0.44 & 0.03 & 0.39 & 0.25 & 0.05 & 0.10 & 0.07 \\
   &    & Qwen 3 (4B) & 0.31 & 0.49 & 0.40 & \underline{\textbf{0.42}} & 0.60 & 0.52 & 0.13 & 0.24 & 0.19 & 0.27 & 0.34 & 0.31 \\
   &    & Qwen 3 (32B) & \underline{\textbf{0.36}} & \underline{\textbf{0.52}} & \underline{\textbf{0.45}} & 0.42 & \lrg 0.51 & 0.47 & 0.15 & 0.29 & 0.23 & \underline{\textbf{0.31}} & \underline{\textbf{0.36}} & \underline{\textbf{0.34}} \\
   &    & Llama 3.3 & 0.35 & 0.46 & 0.41 & 0.40 & \textbf{0.68} & 0.54 & \textbf{0.45} & 0.61 & 0.54 & 0.25 & 0.29 & 0.28 \\
   &    & DeepSeek-R1 & 0.34 & 0.49 & 0.42 & 0.42 & 0.65 & \textbf{0.54} & 0.43 & \textbf{0.67} & \textbf{0.57} & 0.26 & 0.33 & 0.30 \\
   &    & Command R+ & 0.29 & 0.46 & 0.38 & 0.38 & 0.64 & 0.51 & 0.32 & 0.44 & 0.39 & 0.21 & 0.27 & 0.25 \\
   \cmidrule{2-15}
   & \multirow[c]{5}{*}{\react} & Llama 3.2 & 0.11 & \lrg 0.23 & \lrg 0.17 & \lrg 0.17 & \lrg 0.38 & \lrg 0.27 & 0.15 & 0.36 & 0.27 & 0.04 & 0.09 & 0.06 \\
   &    & Qwen 3 (4B) & 0.33 & 0.43 & 0.38 & 0.40 & 0.66 & 0.53 & 0.01 & 0.02 & 0.02 & 0.27 & 0.29 & 0.28 \\
   &    & Qwen 3 (32B) & 0.35 & 0.47 & 0.42 & 0.41 & 0.63 & 0.52 & 0.11 & 0.22 & 0.17 & \textbf{0.29} & \textbf{0.35} & \textbf{0.33} \\
   &    & Llama 3.3 & \textbf{0.36} & \textbf{0.48} & \textbf{0.42} & \textbf{0.42} & \underline{\textbf{0.73}} & \underline{\textbf{0.56}} & \textbf{0.62} & \textbf{0.73} & \textbf{0.68} & 0.27 & 0.31 & 0.29 \\
   &    & Command R+ & 0.30 & 0.44 & 0.37 & 0.35 & \lrg 0.50 & 0.42 & 0.17 & 0.23 & 0.20 & 0.19 & 0.26 & 0.23 \\
   \cmidrule{2-15}
   & \multirow[c]{5}{*}{\pe} & Llama 3.2 & \lrg 0.02 & \lrg 0.05 & \lrg 0.04 & \lrg 0.03 & \lrg 0.11 & \lrg 0.07 & 0.04 & 0.12 & 0.09 & \lrg 0.01 & \lrg 0.01 & \lrg 0.01 \\
   &    & Qwen 3 (4B) & \lrg 0.04 & \lrg 0.07 & \lrg 0.05 & \lrg 0.06 & \lrg 0.11 & \lrg 0.08 & 0.01 & 0.02 & 0.02 & 0.03 & \lrg 0.03 & \lrg 0.03 \\
   &    & Qwen 3 (32B) & 0.29 & \textbf{0.46} & 0.38 & \textbf{0.38} & \lrg 0.58 & 0.48 & 0.11 & 0.37 & 0.27 & \textbf{0.24} & \textbf{0.3} & \textbf{0.27} \\
   &    & Llama 3.3 & \textbf{0.31} & \textbf{0.46} & \textbf{0.39} & 0.37 & \textbf{0.60} & \textbf{0.49} & \underline{\textbf{0.72}} & \underline{\textbf{0.78}} & \underline{\textbf{0.75}} & 0.23 & 0.29 & 0.26 \\
   &    & Command R+ & 0.23 & 0.36 & 0.30 & 0.29 & \lrg 0.52 & 0.41 & 0.18 & 0.27 & 0.23 & 0.14 & 0.20 & 0.17 \\
   \cmidrule{2-15}
   & \random &  & 0.10 & 0.30 & 0.20 & 0.20 & 0.60 & 0.40 & - & - & - & 0.02 & 0.06 & 0.04 \\
\midrule
\multirow[c]{17}{*}{$\mathcal{B}$} & \multirow[c]{6}{*}{\baselinetab} & Llama 3.2 & 0.03 & 0.10 & 0.06 & \lrg 0.06 & \lrg 0.17 & \lrg 0.12 & 0.01 & 0.04 & 0.02 & \lrg 0.0 & 0.007 & 0.003 \\
   &    & Qwen 3 (4B) & 0.07 & 0.19 & 0.13 & 0.07 & \lrg 0.19 & 0.13 & 0.14 & 0.37 & 0.28 & 0.015 & 0.041 & 0.029  \\
   &    & Qwen 3 (32B) & \textbf{0.07} & \textbf{0.22} & \textbf{0.14} & 0.08 & \textbf{0.24} & 0.16 & 0.18 & 0.34 & 0.28 & \textbf{0.019} & \textbf{0.074} & \textbf{0.044} \\
   &    & Llama 3.3 & 0.03 & 0.16 & 0.10 & \textbf{0.09} & 0.24 & \textbf{0.16} & 0.10 & 0.45 & 0.30 & 0.005 & 0.020 & 0.013 \\
   &    & DeepSeek-R1 & 0.03 & 0.16 & 0.10 & \lrg 0.06 & \lrg 0.19 & \lrg 0.12 & 0.19 & \textbf{0.59} & \textbf{0.41} & 0.007 & 0.025 & 0.016 \\
   &    & Command R+ & 0.03 & 0.08 & 0.06 & 0.07 & \lrg 0.19 & \lrg 0.13 & \textbf{0.28} & 0.47 & 0.38 & 0.005 & 0.019 & 0.012 \\
   \cmidrule{2-15}
   & \multirow[c]{5}{*}{\react} & Llama 3.2 & 0.02 & 0.07 & 0.05 & \lrg 0.03 & \lrg 0.07 & \lrg 0.05 & 0.21 & 0.25 & 0.23 & \lrg 0.0 & \lrg 0.003 & \lrg 0.002 \\
   &    & Qwen 3 (4B) & \underline{\textbf{0.11}} & \underline{\textbf{0.33}} & \underline{\textbf{0.23}} & 0.08 & 0.22 & 0.15 & 0.01 & 0.08 & 0.05 &  0.024 & 0.083 & 0.059 \\
   &    & Qwen 3 (32B) & 0.10 & 0.24 & 0.17 & \underline{\textbf{0.10}} & \underline{\textbf{0.25}} & \underline{\textbf{0.18}} & 0.20 & 0.30 & 0.26 & 
   \underline{\textbf{0.035}} & \underline{\textbf{0.088}} & \underline{\textbf{0.062}} \\
   &    & Llama 3.3 & 0.04 & 0.14 & 0.10 & \lrg 0.05 & \lrg 0.17 & \lrg 0.11 & \textbf{0.54} & \underline{\textbf{0.73}} & \underline{\textbf{0.65}} & 0.003 & 0.017 & 0.010 \\
   &    & Command R+ & 0.06 & 0.10 & 0.08 & \lrg 0.06 & \lrg 0.12 & \lrg 0.09 & 0.26 & 0.32 & 0.29 & 0.008 & 0.015 & 0.012 \\
   \cmidrule{2-15}
   & \multirow[c]{5}{*}{\pe} & Llama 3.2 & \lrg 0.01 & \lrg 0.03 & \lrg 0.02 & \lrg 0.01 & \lrg 0.03 & \lrg 0.02 & 0.03 & 0.04 & 0.03 & 0.002 & \lrg 0.003 & 0.003 \\
   &    & Qwen 3 (4B) & \lrg 0.01 & \lrg 0.03 & \lrg 0.02 & \lrg 0.01 & \lrg 0.02 & \lrg 0.01 & 0.0 & 0.01 & 0.01 & \lrg 0.0 & \lrg 0.003 & \lrg 0.002 \\
   &    & Qwen 3 (32B) & \textbf{0.09} & \textbf{0.24} & \textbf{0.17} & \textbf{0.08} & \textbf{0.21} & \textbf{0.15} & 0.23 & 0.35 & 0.30 & \textbf{0.024} & \textbf{0.061} & \textbf{0.042} \\
   &    & Llama 3.3 & 0.02 & 0.11 & 0.06 & \lrg 0.05 & \lrg 0.13 & \lrg 0.09 & \underline{\textbf{0.54}} & \textbf{0.59} & \textbf{0.57} & 0.002 & 0.014 & 0.007 \\
   &    & Command R+ & 0.03 & 0.07 & 0.06 & \lrg 0.07 & \lrg 0.13 & \lrg 0.10 & 0.22 & 0.27 & 0.24 & 0.008 & 0.014 & 0.011 \\
   \cmidrule{2-15}
   & \random &  & 0.02 & 0.05 & 0.04 & 0.07 & 0.20 & 0.13 & - & - & - & 0.001 & 0.004 & 0.002 \\
\bottomrule
\end{tabular}
\par\smallskip
\begin{minipage}{0.91\linewidth}
    \footnotesize
    Per metric column: the top value within each dataset–workflow is shown in \textbf{bold}; the top value across workflows (per dataset) is \underline{underlined}; values below the random-guessing baseline are shown in \lrgtext{color}.
\end{minipage}
\end{table*}


{\looseness=-1
We report the accuracy of the root-cause locations (LA), types (TA), propagation paths (PA), and overall hypotheses (HA) in \Cref{tab:accuracy-results}.
We include a random-guessing baseline for each dataset
(computed using the number of fault locations and fault type categories for LA and TA, respectively).
\B~represents a substantially harder problem than \A due to the larger hypothesis space
(see \Cref{tab:dataset-details}). 
\par}

{\looseness=-1
\observation{No one-size-fits-all---model strengths are task-specific.}
Table~\ref{tab:accuracy-results} shows Qwen 3 (32B) consistently attains the highest HA across all workflows and datasets,
with HA@3 up to 0.36 on \A~(\baseline) and 0.09 on \B~(\react).
Qwen 3 (32B) and Llama 3.3 exhibit similar LA and TA performance (2.6$\pm$4.9 difference).
However, Qwen 3 (32B) leads on \pe~in \B, while Llama 3.3 achieves markedly higher PA across all workflows. 
Surprisingly, Qwen 3 (4B) performs competitively with Qwen 3 (32B) for \baseline~and \react, averaging only 1.2$\pm$3.5 accuracy points lower (max 8.8), excluding PA for \react~which does not follow this trend.
Taken together, these results show a task-specific specialization: Qwen~3~(32B) is the best single-model choice for overall hypothesis synthesis---its 4B counterpart is a suitable choice for smaller systems or less complex agentic workflows---while Llama~3.3 is preferable when path reconstruction and grounding in the system context is the objective.
No single model dominates across all RCA subtasks. Model selection should therefore reflect the target RCA objective (e.g., ranking hypotheses, path reconstruction) rather than model size alone.
\par}


{\looseness=-1
\observation{Agentic workflows show diminishing or negative returns on LA, TA, and HA, and increase below random-guessing rates in specific cases.}
Across many configurations, \baseline~often equals or outperforms \react~and \pe~for these metrics.
For the strongest models (Qwen~3~(32B), Llama~3.3), differences between their respective \react~and \pe~results are typically small (within a few percentage points), indicating limited marginal gains from the added agentic structure.
By contrast, smaller models degrade substantially with increased workflow complexity, e.g., Llama~3.2 LA@3 falls from 0.31 (\baseline) to 0.23 (\react) and to 0.05 (\pe). 
Crucially, rates of below random-guessing performance rise with workflow complexity: TA is below its random baseline for many model/workflow pairs, while LA and HA fall below random primarily for smaller models (Llama~3.2, Qwen~3~(4B)) as workflows become more agentic.
Agentic workflows increase the cognitive and procedural demand on the model. When model capacity is insufficient, added structure does not guide reasoning; rather, it amplifies compounding missteps, producing systematic errors.
\par}

{\looseness=-1
\observation{Intentional exploration via agentic workflows improves PA.}
For sufficiently capable models, intentional exploration via agentic workflows improves propagation path validity: e.g., Llama~3.3 shows PA@3 increases from 0.61 to 0.73 (\baseline~to \react) and up to 0.78 (in \pe) on \A. 
Agentic workflows can better ground their responses in the system knowledge (i.e., KG) and trace plausible propagation paths.
However, we note that this benefit is conditional on the model's ability to reliably plan and execute tool-based exploration.
\par}

\observation{Execution errors and limited tool coverage severely constrain end-to-end utility for small models.}
Practical gains are curtailed by execution errors (graph recursion limits, replanning errors) concentrated in small models: Qwen 3 (4B) and Llama~3.2 suffer high failure rates under \pe~(87.3\% and 77.3\%, respectively). 
Qwen 3 (4B) also shows poor PA under \react~and \pe: it explored the KG with tools in only 23.4\% of \react~samples (19.8\% of non-error samples) and 49.5\% of \pe~samples (67.8\% of non-error samples), versus $\sim$94.5\% tool coverage on average for other models. 
Omitted tool calls force the model to guess or hallucinate paths.
Llama 3.2 is similarly impacted under \pe~(though less so under \react).
Execution errors reflect a capacity mismatch---the iterative planning depth and attention to previous steps required by \pe~exceed what these smaller models can sustain, producing systematic execution breakdowns rather than benign uncertainty.
This means end-to-end performance and real-world utility remain severely limited in agentic workflows.

\conclusionbox{
\textbf{\underline{Answer to \rqone}}:
We find that the RCA task still proves to be challenging for competitive open-source LLMs, with sub-guessing performance arising as workflow complexity scales, particularly for smaller models. 
Although agentic workflows can improve PA, suggesting better groundedness in the system context, it is highly model-specific whether LLMs can adequately exploit agentic workflows. We observe that the utility of smaller models in more complex agent workflows like \pe~is critically constrained by high recursion limit failures, which reflect a capacity mismatch: the multi-step planning these workflows demand exceeds what smaller models can sustain.
}

\vspace{-10pt}
\subsection{\rqtwo: \rqii} \label{sec:results-rq2}

\begin{figure}[!tb]
    \centering
    \includegraphics[width=\linewidth]{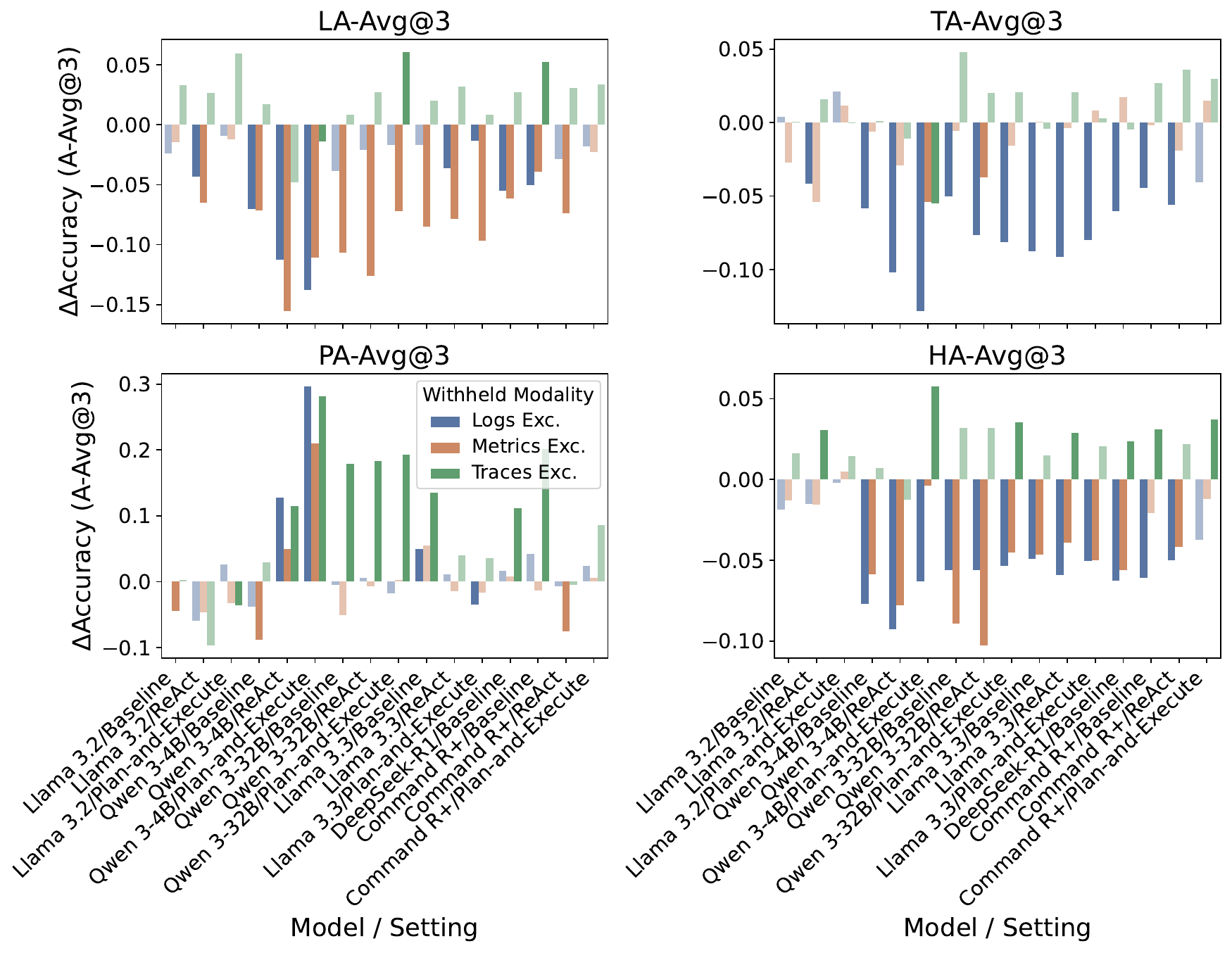} 
    \caption{Change in accuracy for withheld alert modalities. Solid bars denote statistically significant differences ($p < 0.05$) according to the Wilcoxon signed-rank test, while faded bars indicate changes that are not statistically significant.}
    \label{fig:accuracy-withheld-modalities}
    \vspace{-4pt}
\end{figure}

\begin{figure*}[t]
    \centering
    \includegraphics[width=0.99\textwidth]{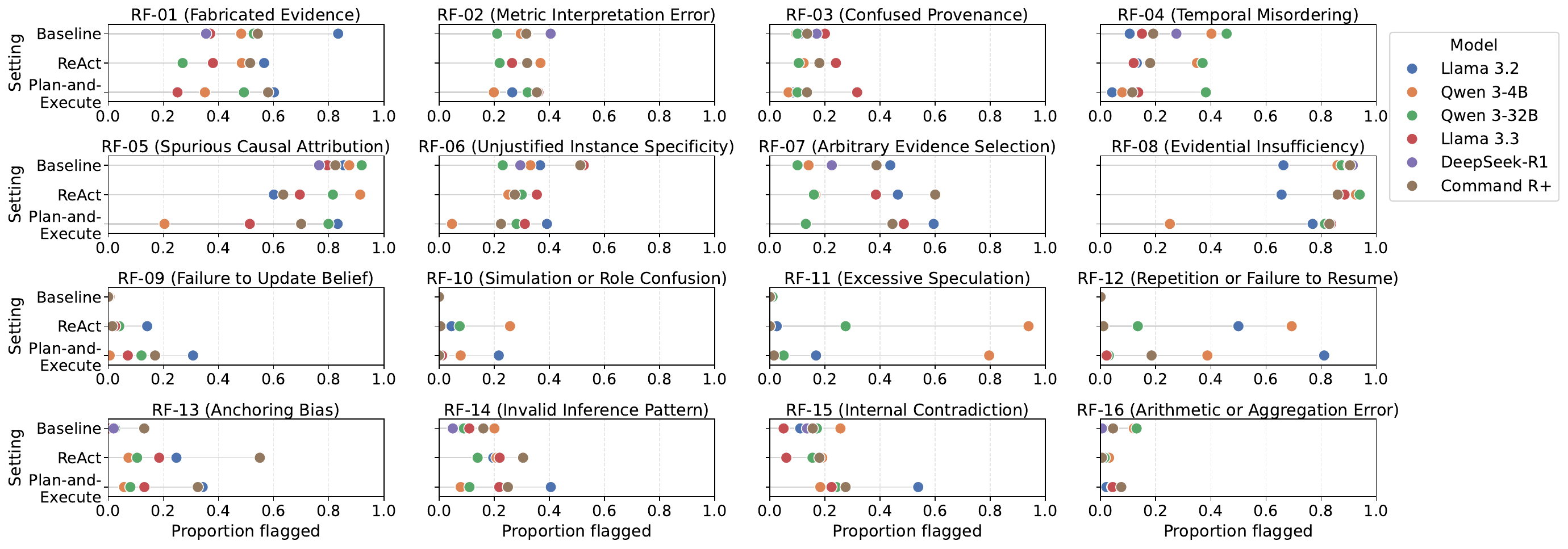} 
    \caption{Prevalence of reasoning failures (RF) in RCA outputs.}
    \label{fig:rf-proportions}
\end{figure*}

To answer \rqtwo, we conduct holdout experiments wherein we withhold each data modality during the alert unification procedure, thereby separately excluding traces, logs, and metrics during RCA inference.
\Cref{fig:accuracy-withheld-modalities} shows the change in accuracy ($\Delta \text{A}$) for each withheld modality aggregated across both datasets.
We apply the Wilcoxon signed-rank test~\cite{Wilcoxon:JSTOR:1945} for statistical significance on the per-sample correctness scores.
The per-sample score is the mean of A@1, A@2, and A@3. It assigns a value of 1.00, 0.67, 0.33, or 0.00 depending on whether the correct root-cause value was found at rank 1, 2, 3, or not found, respectively.
We observe that different modalities affect RCA performance in distinct ways.

\observation{Withholding metrics strongly degrades localization.}
Across nearly all models and workflows, excluding metrics produces the largest and most consistent drop in localization ($\Delta \text{LA}$ typically between --0.07 and --0.15).
For example, Qwen 3 (4B) shows $\Delta\text{LA}=-0.155$.
This indicates metric anomalies are the modality most effectively leveraged by the LLM reasoners for identifying root-cause locations.

\observation{Withholding logs primarily harms fault-type attribution.}
Excluding logs produces the largest degradation in TA ($\Delta \text{TA}$ down to --0.13).
E.g., Qwen 3 (4B) under \pe~has $\Delta\text{TA}=-0.128$.
This suggests that textual alerts contain semantic cues that models use to infer error type.

{\looseness=-1
\observation{Traces often distract; removing them improves accuracy}.
Surprisingly, excluding traces typically increases 
PA (up to +0.28) and HA (up to +0.06).
Qualitatively, we found that LLMs often focus closely on call relationships from trace alerts and ignore other possible entity relationships (e.g., deployment). Additionally, trace alerts were often noisy and voluminous, making alert triage more difficult and drowning out important yet infrequent signals.
In their absence, we noticed that the analysis and exploration of the LLM were more focused and better grounded in the system KG.
\par}

{\looseness=-1
These effects compound in the overall HA:
trace exclusion produces the strongest positive shift ($\Delta\text{HA}$ up to +0.07), while log and metric exclusions yield degradation ($\Delta\text{HA}$ down to --0.10).
\par}

\conclusionbox{
\textbf{\underline{Answer to \rqtwo}}:
Excluding metrics or logs markedly reduces accuracy, indicating that these modalities are most effectively leveraged by LLMs---metrics for fault localization and logs for interpreting fault type.
Conversely, excluding traces often improves accuracy, suggesting that current models struggle to integrate raw trace alerts and that their inclusion can distract reasoning.
Overall, these results identify which modalities LLMs can reason with effectively and which require better integration strategies for LLM-based RCA.
}

\vspace{-6pt}
\subsection{\rqthree: \rqiii} \label{sec:results-rq3}

To answer \rqthree, we identify common reasoning failures observed in RCA inference traces (\Cref{tab:reasoning-failures-taxonomy}) and quantify their prevalence, following the procedure described in \Cref{sec:reasoning-quality}.
\Cref{fig:rf-proportions} shows the relative proportions of RFs identified by the LLM-as-a-Judge.

\observation{
Procedural reasoning failures vary substantially across models and workflows, indicating model-specific instability under agentic control.
}
Procedural RFs---including failure to update belief (RF-09), simulation/role confusion (RF-10), excessive speculation (RF-11), and repetition/stalling (RF-12)---show the widest spread of RF prevalence across models and workflows. 
Numeric/aggregation errors (RF-16) are rare across all conditions.
For agentic workflows, 
Llama 3.2 and Qwen 3 (4B, 32B) commit procedural errors at a far higher frequency.
Prevalence rises from \react~to \pe~for Llama 3.2, while decreasing for Qwen 3 (4B, 32B), indicating differing sensitivity to iterative planning.
This suggests that agentic workflows can amplify latent procedural weaknesses in some models while stabilizing reasoning in others, underscoring the need for model-specific workflow tuning rather than treating planning as universally beneficial.
Prevalence for other (larger) models is comparatively low. 
We hypothesize that small models struggle to exploit agentic procedures.

\observation{
RCA-specific failures are consistent across models, reflecting domain-grounding rather than workflow sensitivity.
}
Metric interpretation errors (RF-02), provenance confusion (RF-03), incorrect instance-specificity (RF-06), and arbitrary triage (RF-07) show few clear trends across models---Qwen models (4B, 32B) are at the lower end for most prevalence counts---and minimal differences between \react~and \pe.
The comparatively narrow spread across models suggests that these failures stem from the inability to perform RCA-specific reasoning, indicating insufficient domain specification.
We suggest that improving RCA performance will likely require explicit domain guidance to properly interpret, process, and prioritize the input alert information, and reasoning patterns (e.g., triaging heuristics, common propagation mechanisms) that an LLM should follow. Fine-tuning on these behaviors could also be beneficial.

\observation{
General per-hypothesis reasoning failures are widespread, indicating that hypothesis claims are often weakly supported.
}
This subset (fabricated evidence (RF-01), evidential insufficiency (RF-08), temporal misordering (RF-04), spurious causality (RF-05)) has the highest average prevalence across RFs, meaning that single-hypothesis statements are frequently under-justified or logically inconsistent.
Notably, RF-04 appears more often in Qwen models.
The high prevalence suggests that hypotheses should not be trusted at face value; effective RCA systems must apply evidence sufficiency checks, temporal ordering validators, and KG-consistency filters before ranking or selecting hypotheses.

{\looseness=-1
\observation{
General cross-cutting reasoning failures vary by model family, with ``thinking''-oriented models showing lower prevalence under agentic workflows.
}
Llama 3.2 and Command R+ show consistently higher prevalence of cross-cutting RFs (e.g., anchoring, internal contradiction, invalid inference patterns), particularly in multi-step traces.
In contrast, Qwen models exhibit lower rates of these failures under both \react~and \pe.
This suggests that global reasoning coherence may depend more on model training style and inductive biases than workflow structure alone.
\par}

\conclusionbox{
\textbf{\underline{Answer to \rqthree}}:
RFs are pervasive across models and workflows and group into \textit{procedural}, \textit{RCA-specific}, and \textit{general} failure categories. 
Prevalence varies by model and workflow (notably elevated procedural RFs in Llama 3.2 and Qwen 3 variants under agentic control), and general per-hypothesis failures have the highest average prevalence. RCA-specific reasoning remains insufficient, and general (non-domain-specific) RFs continue to dominate outputs.
}

\vspace{-6pt}
\subsection{\rqfour: \rqiv} \label{sec:results-rq4}

\begin{figure}[!tb]
    \centering
    \includegraphics[width=\linewidth]{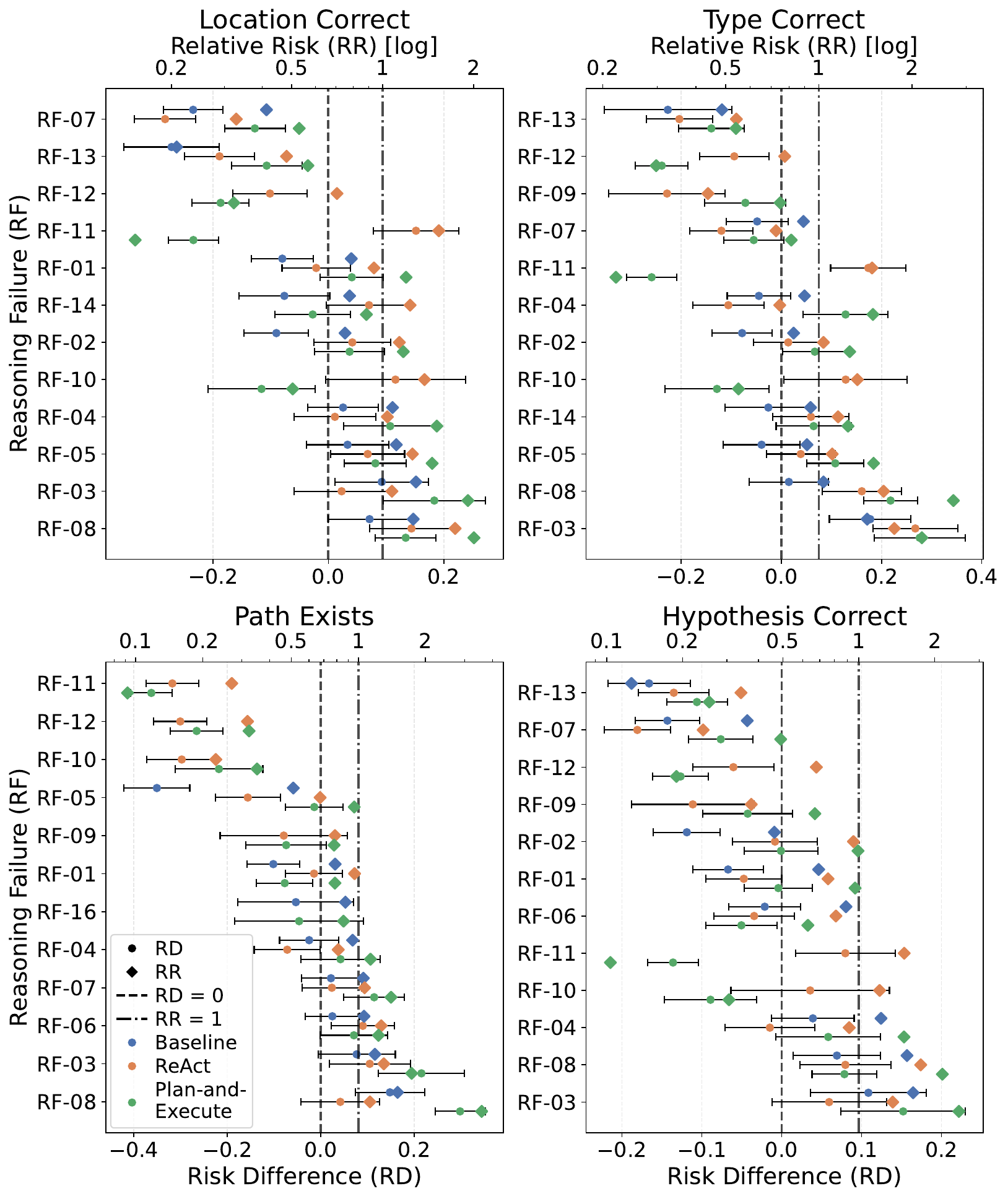} 
    \caption{
    Risk difference (RD, Wilson 95\% CI) and relative risk (RR, log scale) for the top-12 RFs. 
    }
    \label{fig:rf-risk}
    \vspace{-2mm}
\end{figure}

To answer \rqfour, we quantify the association between detected RFs (by the LLM-as-a-Judge) and hypothesis correctness (generated during RCA Inference) using two measures: 
\textit{risk difference} 
$\text{RD} = p_1 - p_0$ 
and \textit{relative risk} 
$\text{RR} = p_1 / p_0$,
where $ p_1 = P(\text{correct} | \text{RF present})$ and $p_0 = P(\text{correct} | \text{RF absent})$.
\Cref{fig:rf-risk} summarizes results across correctness types.
By convention, RD<0 (or RR<1) indicates an RF associated with reduced correctness,
while RD>0 (or RR>1) indicates an association with improved correctness.

\observation{
The clearest negative predictors of correctness are anchoring bias (RF-13), repetition or stalled progress (RF-12), arbitrary evidence selection (RF-07), and failure to update belief (RF-09).
}
Across location, type, and hypothesis, these RFs correspond to at least a 15\% drop in correctness (RD<--0.15) and make correct predictions at least 45\% less likely than when the failure is absent (RR<0.55).
For path existence, a distinct set---simulation or role confusion (RF-10), excessive speculation (RF-11), and RF-12---shows the strongest negative associations (RD<--0.29; RR<0.32), suggesting cases where reasoning becomes decoupled from the system knowledge grounding.

\observation{
Agentic workflows tend to be associated with smaller absolute risk differences (RDs closer to zero) than the non-agentic \baseline~across metrics.
}
This pattern indicates that correct outcomes are less strongly associated with the presence of RFs under agentic workflows, consistent with the view that iterative reasoning and answer formulation may dilute the marginal influence of individual reasoning failures.
Notably, \pe~often exhibits significantly higher RD values (closer to zero) than \baseline. 
However, agentic traces contain more RFs on average (\react~5.1; \pe~4.8; \baseline~4.6). RF co-occurrence (multiple RFs in one trace) and early error compounding (an RF that cascades) could contribute to the observation from \rqone~(agentic workflows do not reliably improve accuracy). Thus, although per-RF associations are weaker under agentic workflows, compounding effects can produce net accuracy loss.

\observation{
The relative association between RFs and correctness differs substantially between \react~and \pe.
}
In most cases, \pe~shows higher (less negative) RD and RR values (though not always statistically significant), but two exceptions stand out.
Repetition or failure to resume (RF-12) is more strongly associated with reduced correctness under \pe~than under \react~for location, type, and hypothesis correctness ($\Delta \text{RR} \approx 0.38$), suggesting that the replanning step may correlate with persistence of stalled reasoning rather than its resolution, even when prior steps and actions are available.
A more extreme divergence occurs for RF-10 and RF-11, which appear positively associated with correctness under \react~(RR>1.49) but negatively under \pe~(RR<0.22).
These RFs co-occur frequently with execution errors (e.g., recursion limit or replan errors) in 68.6\%/80.6\% of \pe~versus 7.4\%/0.5\% of \react~samples for RF-10/RF-11, respectively.
This pattern reflects the frequent co-occurrence of speculative or fabricated assumptions and role-confused reasoning with workflow errors in \pe.
Regardless of workflow, both RF-10 and RF-11 show strong negative associations with path recovery (RD<--0.29; RR<0.32), consistent with loss of grounding in the system knowledge (KG).

A small subset of RFs---notably confused provenance (RF-03) and evidential insufficiency (RF-08)---exhibit positive RD (RR>1), an ostensibly counterintuitive outcome for failure labels. 
This arises from our per-sample aggregation scheme: correctness is marked if \textit{any} attribute is correct, and per-hypothesis RFs are flagged if they occur in \textit{any} hypothesis within the sample.
Consequently, an RF appearing in one hypothesis can co-occur with a different correct hypothesis in the same sample, producing an apparent net positive association. 
Qualitatively, we observed that models often ``reach" for additional (second/third) hypotheses (as required by the task) when such RFs are flagged. This artifact reflects the forced three-hypothesis task (even if confident in only one or two).

\conclusionbox{
\textbf{\underline{Answer to \rqfour}:}
The presence of reasoning failures---particularly anchoring bias (RF-13), repetition/stalled progress (RF-12), arbitrary evidence selection (RF-07), and failure to update belief (RF-09)---is associated with substantially reduced RCA correctness.
Agentic workflows are often associated with smaller negative risk differences, but do not eliminate them. However, RF co-occurrence and early failure compounding remain plausible contributors to the accuracy degradations observed in \rqone.
}
\vspace{-12pt}
\section{Limitations and Threats to Validity}

\noindent
\textbf{Limitations.}
{\looseness=-1
Our study adopts several controlled simplifications to isolate and evaluate reasoning, which naturally introduce limitations.
First, the alerts provided to the agent are due to a \textit{single} known root-cause fault. In practice, concurrent faults can occur, producing overlapping and interleaved alerts that require alert correlation and attribution to disambiguate causality. Since our goal is to isolate reasoning in the RCA task, we therefore restrict to single-fault scenarios.
Although we follow practices common in prior work for alert extraction, the alerts provided to the agent are not gold-labeled and may contain noise. 
However, imperfect monitoring and signal quality remain a realistic concern in real-world production systems.
Optimized alert selection, prioritization, or representation 
may help address this in future work.
\par}



\noindent
\textbf{Internal Validity.}
{\looseness=-1
To isolate and assess the reasoning behavior of LLM agents, we minimize confounding factors through explicit natural-language prompting and data representation, and by reducing the action space to deterministic tools with clearly defined semantics. However, results may still be sensitive to the phrasing of prompts, tool descriptions, or representations of system information (e.g., alerts, KG), which can subtly influence agent behavior.
We mitigate some of these effects by incorporating different representation strategies of system information.
%
Taxonomy elicitation relies on qualitative coding and human interpretation, which can introduce subjectivity. Structured definitions, exemplars, and paired intermediate LLM–human annotation rounds were used to align interpretations, but some coder bias may remain.
The taxonomy is also induced from a limited number of samples (170), which may not capture rare reasoning failure types; however, we believe it adequately captures common and impactful failures. 
Finally, LLM-based judging introduces uncertainty due to model preferences and inductive biases. We mitigate this by employing a distinct, strong evaluator model (GPT-5), adhering to best practices for LLM-as-a-Judge systems, and conducting an independent two-author review of a random subset of annotations to increase confidence in the reliability of our results.
\par}

\noindent
\textbf{External Validity.}
{\looseness=-1
Our evaluation is limited to a small set of open-source benchmarks and LLM models. While these case studies represent diverse and realistic microservice-based systems and we consider LLM models from a variety of model families, our findings may not generalize to other domains, large-scale production environments, or proprietary models. 
Moreover, our findings are specific to the agent workflows and input modalities explored in this work.
To mitigate this threat, we make our methodology and evaluation strategy publicly available~\cite{appendix} to encourage replication and extension studies
to further assess the generalizability of the findings.
\par}

\vspace{-3pt}
\section{Conclusion} \label{sec:conclusion}

{\looseness=-1
In this work, we present a reasoning-centered evaluation framework for LLM-based RCA agents that isolates the reasoning process from framework complexity. 
By simplifying the RCA setup, our approach enables transparent assessment of LLM reasoning capabilities for RCA.
We find that the simplified setup still poses ample challenge: accuracy is low across models and workflows, and reasoning failures are widespread. Agentic workflows often yield diminishing or negative returns, particularly for smaller models. 
Yet, interactive exploration remains necessary---providing full system context upfront is infeasible as systems scale in production environments. This tension suggests that progress will require improving specific reasoning capabilities rather than relying on agentic structure alone.
\par}

Our analysis shows that RCA-specific reasoning by LLMs is frequently inconsistent with standard diagnostic heuristics and causal propagation mechanisms, while general and procedural failures (e.g., anchoring, arbitrary evidence selection, stalled reasoning) strongly predict incorrect outcomes. 
Addressing these will likely require encouraging early hypothesis diversification (e.g., via tree-of-thought), self-consistency or critique mechanisms, evidence sufficiency checks, and explicit domain guidance for triage and causal reasoning.
Finally, understanding how reasoning failures co-occur and compound remains an important open direction, particularly in multi-step or multi-agent settings where early missteps cascade.
Overall, our findings highlight the need for transparency and reasoning quality in RCA agent design.
Our reasoning failure taxonomy and rationale-based evaluation scheme offer reusable components for diagnosing and improving RCA reasoning.
More broadly, this work moves toward RCA agents that are not only automated but also interpretable, inspectable, and better aligned with the practical demands of fault diagnosis in complex systems.

\vspace{-1pt}
\begin{acks}
%
Supported by General Motors of Canada Company, with thanks to
Abhishek Kumar Dubey and S. Ramesh 
for their fruitful discussion.

\end{acks}

\clearpage
\bibliographystyle{ACM-Reference-Format}
\bibliography{12-bibliography}

@article{Zhang:DiagFusion:TSC:2023,
  author  = {Zhang, Shenglin and Jin, Pengxiang and Lin, Zihan and Sun, Yongqian and Zhang, Bicheng and Xia, Sibo and Li, Zhengdan and Zhong, Zhenyu and Ma, Minghua and Jin, Wa and Zhang, Dai and Zhu, Zhenyu andlas Pei, Dan},
  journal = {IEEE Transactions on Services Computing},
  title   = {{Robust Failure Diagnosis of Microservice System Through Multimodal Data}},
  year    = {2023},
  volume  = {16},
  number  = {6},
  pages   = {3851-3864},
  doi     = {10.1109/TSC.2023.3290018}
}

@inproceedings{Zheng:MULAN:WWW:2024,
  author    = {Zheng, Lecheng and Chen, Zhengzhang and He, Jingrui and Chen, Haifeng},
  title     = {{MULAN: Multi-modal Causal Structure Learning and Root Cause Analysis for Microservice Systems}},
  year      = {2024},
  isbn      = {9798400701719},
  publisher = {Association for Computing Machinery},
  address   = {New York, NY, USA},
  doi       = {10.1145/3589334.3645442},
  booktitle = {Proceedings of the ACM Web Conference (WWW)},
  pages     = {4107–4116},
  numpages  = {10},
  location  = {Singapore, Singapore},
}

@inproceedings{Guangba:Nezha:FSE:2023,
  author    = {Yu, Guangba and Chen, Pengfei and Li, Yufeng and Chen, Hongyang and Li, Xiaoyun and Zheng, Zibin},
  title     = {{Nezha: Interpretable Fine-Grained Root Causes Analysis for Microservices on Multi-modal Observability Data}},
  year      = {2023},
  isbn      = {9798400703270},
  publisher = {Association for Computing Machinery},
  address   = {New York, NY, USA},
  url       = {https://doi.org/10.1145/3611643.3616249},
  doi       = {10.1145/3611643.3616249},
  booktitle = {Proceedings of the 31st ACM Joint European Software Engineering Conference and Symposium on the Foundations of Software Engineering (ESEC/FSE)},
  pages     = {553–565},
  numpages  = {13},
  location  = {San Francisco, CA, USA},
}

@inproceedings{Lee:Eadro:ICSE:2023,
  author    = {Lee, Cheryl and Yang, Tianyi and Chen, Zhuangbin and Su, Yuxin and Lyu, Michael R.},
  booktitle = {IEEE/ACM 45th International Conference on Software Engineering (ICSE)},
  title     = {{Eadro: An End-to-End Troubleshooting Framework for Microservices on Multi-source Data}},
  year      = {2023},
  volume    = {},
  number    = {},
  pages     = {1750-1762},
  doi       = {10.1109/ICSE48619.2023.00150}
}

@ARTICLE{Han:HolisticRCA:TSC:2024,
  author={Han, Yongqi and Du, Qingfeng and Huang, Ying and Li, Pengsheng and Shi, Xiaonan and Wu, Jiaqi and Fang, Pei and Tian, Fulong and He, Cheng},
  journal={IEEE Transactions on Services Computing}, 
  title={{Holistic Root Cause Analysis for Failures in Cloud-Native Systems Through Observability Data}}, 
  year={2024},
  volume={17},
  number={6},
  pages={3789-3802},
  doi={10.1109/TSC.2024.3478759}
}

@inproceedings{Ahmed:LLM-Finetune:ICSE:2023,
  author    = {Ahmed, Toufique and Ghosh, Supriyo and Bansal, Chetan and Zimmermann, Thomas and Zhang, Xuchao and Rajmohan, Saravan},
  booktitle = {IEEE/ACM 45th International Conference on Software Engineering (ICSE)},
  title     = {{Recommending Root-Cause and Mitigation Steps for Cloud Incidents using Large Language Models}},
  year      = {2023},
  volume    = {},
  number    = {},
  pages     = {1737-1749},
  doi       = {10.1109/ICSE48619.2023.00149}
}

@inproceedings{Wang:RCAGent:CIKM:2024,
    author = {Wang, Zefan and Liu, Zichuan and Zhang, Yingying and Zhong, Aoxiao and Wang, Jihong and Yin, Fengbin and Fan, Lunting and Wu, Lingfei and Wen, Qingsong},
    title = {{RCAgent: Cloud Root Cause Analysis by Autonomous Agents with Tool-Augmented Large Language Models}},
    year = {2024},
    isbn = {9798400704369},
    publisher = {Association for Computing Machinery},
    address = {New York, NY, USA},
    doi = {10.1145/3627673.3680016},
    booktitle = {Proceedings of the 33rd ACM International Conference on Information and Knowledge Management (CIKM)},
    pages = {4966–4974},
    numpages = {9},
    location = {Boise, ID, USA},
}

@inproceedings{Zhang:LLM-InContext:FSE:2024,
  author    = {Zhang, Xuchao and Ghosh, Supriyo and Bansal, Chetan and Wang, Rujia and Ma, Minghua and Kang, Yu and Rajmohan, Saravan},
  title     = {{Automated Root Causing of Cloud Incidents using In-Context Learning with GPT-4}},
  year      = {2024},
  isbn      = {9798400706585},
  publisher = {Association for Computing Machinery},
  address   = {New York, NY, USA},
  doi       = {10.1145/3663529.3663846},
  booktitle = {Companion Proceedings of the 32nd ACM International Conference on the Foundations of Software Engineering (FSE)},
  pages     = {266–277},
  numpages  = {12},
  location  = {Porto de Galinhas, Brazil},
}

@inproceedings{Roy:LLM-React:FSE:2024,
  author    = {Roy, Devjeet and Zhang, Xuchao and Bhave, Rashi and Bansal, Chetan and Las-Casas, Pedro and Fonseca, Rodrigo and Rajmohan, Saravan},
  title     = {{Exploring LLM-Based Agents for Root Cause Analysis}},
  year      = {2024},
  isbn      = {9798400706585},
  publisher = {Association for Computing Machinery},
  address   = {New York, NY, USA},
  doi       = {10.1145/3663529.3663841},
  booktitle = {Companion Proceedings of the 32nd ACM International Conference on the Foundations of Software Engineering (FSE)},
  pages     = {208–219},
  numpages  = {12},
  location  = {Porto de Galinhas, Brazil},
}

@inproceedings{Zhang:mABC:EMNLP:2024,
    title = {{m{ABC}: Multi-Agent Blockchain-inspired Collaboration for Root Cause Analysis in Micro-Services Architecture}},
    author = "Zhang, Wei  and
      Guo, Hongcheng  and
      Yang, Jian  and
      Tian, Zhoujin  and
      Zhang, Yi  and
      Chaoran, Yan  and
      Li, Zhoujun  and
      Li, Tongliang  and
      Shi, Xu  and
      Zheng, Liangfan  and
      Zhang, Bo",
    editor = "Al-Onaizan, Yaser  and
      Bansal, Mohit  and
      Chen, Yun-Nung",
    booktitle = "Findings of the Association for Computational Linguistics (EMNLP)",
    month = nov,
    year = "2024",
    address = "Miami, Florida, USA",
    publisher = "Association for Computational Linguistics",
    doi = "10.18653/v1/2024.findings-emnlp.232",
    pages = "4017--4033",
}

@inproceedings{Chen:RCACopilot:EuroSys:2024,
  author    = {Chen, Yinfang and Xie, Huaibing and Ma, Minghua and Kang, Yu and Gao, Xin and Shi, Liu and Cao, Yunjie and Gao, Xuedong and Fan, Hao and Wen, Ming and Zeng, Jun and Ghosh, Supriyo and Zhang, Xuchao and Zhang, Chaoyun and Lin, Qingwei and Rajmohan, Saravan and Zhang, Dongmei and Xu, Tianyin},
  title     = {{Automatic Root Cause Analysis via Large Language Models for Cloud Incidents}},
  year      = {2024},
  isbn      = {9798400704376},
  publisher = {Association for Computing Machinery},
  address   = {New York, NY, USA},
  doi       = {10.1145/3627703.3629553},
  booktitle = {Proceedings of the Nineteenth European Conference on Computer Systems (EuroSys)},
  pages     = {674–688},
  numpages  = {15},
  location  = {Athens, Greece},
}

@inproceedings{Goel:LLM-InContext:FSE:2024,
  author    = {Goel, Drishti and Husain, Fiza and Singh, Aditya and Ghosh, Supriyo and Parayil, Anjaly and Bansal, Chetan and Zhang, Xuchao and Rajmohan, Saravan},
  title     = {{X-Lifecycle Learning for Cloud Incident Management using LLMs}},
  year      = {2024},
  isbn      = {9798400706585},
  publisher = {Association for Computing Machinery},
  address   = {New York, NY, USA},
  doi       = {10.1145/3663529.3663861},
  booktitle = {Companion Proceedings of the 32nd ACM International Conference on the Foundations of Software Engineering (FSE)},
  pages     = {417–428},
  numpages  = {12},
  location  = {Porto de Galinhas, Brazil},
}

@ARTICLE{Zhang:TAMO:TSC:2025,
  author={Zhang, Xiao and Wang, Qi and Li, Mingyi and Yuan, Yuan and Xiao, Mengbai and Zhuang, Fuzhen and Yu, Dongxiao},
  journal={IEEE Transactions on Services Computing}, 
  title={{TAMO:Fine-Grained Root Cause Analysis via Tool-Assisted LLM Agent With Multi-Modality Observation Data in Cloud-Native Systems}}, 
  year={2025},
  volume={18},
  number={6},
  pages={4221-4233},
  doi={10.1109/TSC.2025.3629066}
}

@INPROCEEDINGS{Li:COCA:ICSE:2025,
  author={Li, Yichen and Wu, Yulun and Liu, Jinyang and Jiang, Zhihan and Chen, Zhuangbin and Yu, Guangba and Lyu, Michael R.},
  booktitle={IEEE/ACM 47th International Conference on Software Engineering (ICSE)}, 
  title={{COCA: Generative Root Cause Analysis for Distributed Systems with Code Knowledge}}, 
  year={2025},
  volume={},
  number={},
  pages={1346-1358},
  doi={10.1109/ICSE55347.2025.00234}
}

@INPROCEEDINGS{Qiu:RAG-LLM-Hardware:COINS:2025,
  author={Qiu, Siyu and Wang, Muzhi and Afsharmazayejani, Raheel and Shahmiri, Mohammad Moradi and Tan, Benjamin and Pearce, Hammond},
  booktitle={IEEE International Conference on Omni-layer Intelligent Systems (COINS)}, 
  title={{Towards LLM-based Root Cause Analysis of Hardware Design Failures}}, 
  year={2025},
  volume={},
  number={},
  pages={1-6},
  doi={10.1109/COINS65080.2025.11125748}
}

@misc{Xie:TVDiag:arXiv:2025,
      title={{TVDiag: A Task-oriented and View-invariant Failure Diagnosis Framework with Multimodal Data}}, 
      author={Shuaiyu Xie and Jian Wang and Hanbin He and Zhihao Wang and Yuqi Zhao and Neng Zhang and Bing Li},
      year={2025},
      eprint={2407.19711},
      archivePrefix={arXiv},
      primaryClass={cs.SE},
}

@inproceedings{Pei:Flow-of-Action:WWW:2025,
    author = {Pei, Changhua and Wang, Zexin and Liu, Fengrui and Li, Zeyan and Liu, Yang and He, Xiao and Kang, Rong and Zhang, Tieying and Chen, Jianjun and Li, Jianhui and Xie, Gaogang and Pei, Dan},
    title = {{Flow-of-Action: SOP Enhanced LLM-Based Multi-Agent System for Root Cause Analysis}},
    year = {2025},
    isbn = {9798400713316},
    publisher = {Association for Computing Machinery},
    address = {New York, NY, USA},
    doi = {10.1145/3701716.3715225},
    booktitle = {Companion Proceedings of the ACM on Web Conference},
    pages = {422–431},
    numpages = {10},
    location = {Sydney NSW, Australia},
    series = {WWW '25}
}

@inproceedings{Han:LasRCA:ASE:2024,
    author = {Han, Yongqi and Du, Qingfeng and Huang, Ying and Wu, Jiaqi and Tian, Fulong and He, Cheng},
    title = {{The Potential of One-Shot Failure Root Cause Analysis: Collaboration of the Large Language Model and Small Classifier}},
    year = {2024},
    isbn = {9798400712487},
    publisher = {Association for Computing Machinery},
    address = {New York, NY, USA},
    url = {https://doi.org/10.1145/3691620.3695475},
    doi = {10.1145/3691620.3695475},
    booktitle = {Proceedings of the 39th IEEE/ACM International Conference on Automated Software Engineering (ASE)},
    pages = {931–943},
    numpages = {13},
    location = {Sacramento, CA, USA},
}

@inproceedings{Xu:OpenRCA:ICLR:2025,
  title={{OpenRCA: Can Large Language Models Locate the Root Cause of Software Failures?}},
  author={Xu, Junjielong and Zhang, Qinan and Zhong, Zhiqing and He, Shilin and Zhang, Chaoyun and Lin, Qingwei and Pei, Dan and He, Pinjia and Zhang, Dongmei and Zhang, Qi},
  booktitle={The Thirteenth International Conference on Learning Representations (ICLR)},
  year=2025,
  url={https://openreview.net/forum?id=M4qNIzQYpd}
}

@inproceedings{Wang:MRCA:ASE:2024,
    author = {Wang, Yidan and Zhu, Zhouruixing and Fu, Qiuai and Ma, Yuchi and He, Pinjia},
    title = {{MRCA: Metric-level Root Cause Analysis for Microservices via Multi-Modal Data}},
    year = {2024},
    isbn = {9798400712487},
    publisher = {Association for Computing Machinery},
    address = {New York, NY, USA},
    doi = {10.1145/3691620.3695485},
    booktitle = {Proceedings of the 39th IEEE/ACM International Conference on Automated Software Engineering (ASE)},
    pages = {1057–1068},
    numpages = {12},
    location = {Sacramento, CA, USA},
}

@ARTICLE{Sui:LogKG:TSC:2023,
  author={Sui, Yicheng and Zhang, Yuzhe and Sun, Jianjun and Xu, Ting and Zhang, Shenglin and Li, Zhengdan and Sun, Yongqian and Guo, Fangrui and Shen, Junyu and Zhang, Yuzhi and Pei, Dan and Yang, Xiao and Yu, Li},
  journal={IEEE Transactions on Services Computing}, 
  title={{LogKG: Log Failure Diagnosis Through Knowledge Graph}}, 
  year={2023},
  volume={16},
  number={5},
  pages={3493-3507},
  doi={10.1109/TSC.2023.3293890}
}

@misc{Wang:survey:arXiv:2024,
  title         = {{A Comprehensive Survey on Root Cause Analysis in (Micro) Services: Methodologies, Challenges, and Trends}},
  author        = {Tingting Wang and Guilin Qi},
  year          = {2024},
  eprint        = {2408.00803},
  archiveprefix = {arXiv},
  primaryclass  = {cs.SE}
}

@article{Zhang:survey:TSEM:2025,
    author = {Zhang, Shenglin and Xia, Sibo and Fan, Wenzhao and Shi, Binpeng and Xiong, Xiao and Zhong, Zhenyu and Ma, Minghua and Sun, Yongqian and Pei, Dan},
    title = {Failure Diagnosis in Microservice Systems: A Comprehensive Survey and Analysis},
    year = {2025},
    issue_date = {January 2026},
    publisher = {Association for Computing Machinery},
    address = {New York, NY, USA},
    volume = {35},
    number = {1},
    issn = {1049-331X},
    doi = {10.1145/3715005},
    journal = {ACM Trans. Softw. Eng. Methodol.},
    month = dec,
    articleno = {2},
    numpages = {55},
}

@article{Soldani:survey:ACMCS:2022,
  author     = {Soldani, Jacopo and Brogi, Antonio},
  title      = {{Anomaly Detection and Failure Root Cause Analysis in (Micro) Service-Based Cloud Applications: A Survey}},
  year       = {2022},
  issue_date = {March 2023},
  publisher  = {Association for Computing Machinery},
  address    = {New York, NY, USA},
  volume     = {55},
  number     = {3},
  issn       = {0360-0300},
  url        = {https://doi.org/10.1145/3501297},
  doi        = {10.1145/3501297},
  journal    = {ACM Comput. Surv.},
  month      = feb,
  articleno  = {59},
  numpages   = {39}
}

@misc{GAIA:dataset,
  author = {{CloudWise OpenSource}},
  title  = {{GAIA: Generic AIOps Atlas}},
  year   = {2022},
  howpublished = {\url{https://github.com/CloudWise-OpenSource/GAIA-DataSet}},
  note   = {Accessed 2025-06-03}
}

@misc{OnlineBoutique:benchmark,
  author = {Google Cloud Platform},
  title  = {{Online Boutique}},
  howpublished = {\url{https://github.com/GoogleCloudPlatform/microservices-demo}},
  year   = {2020},
  note   = {Accessed 2025-06-03}
}

@inproceedings{Yao:ReAct:ICLR:2023,
    title={{ReAct: Synergizing Reasoning and Acting in Language Models}},
    author={Shunyu Yao and Jeffrey Zhao and Dian Yu and Nan Du and Izhak Shafran and Karthik R Narasimhan and Yuan Cao},
    booktitle={The Eleventh International Conference on Learning Representations (ICLR)},
    year={2023},
    url={https://openreview.net/forum?id=WE_vluYUL-X}
}

@inproceedings{Wei:CoT:NeurIPS:2022,
 author = {Wei, Jason and Wang, Xuezhi and Schuurmans, Dale and Bosma, Maarten and ichter, brian and Xia, Fei and Chi, Ed and Le, Quoc V and Zhou, Denny},
 booktitle = {{Advances in Neural Information Processing Systems (NeurIPS)}},
 editor = {S. Koyejo and S. Mohamed and A. Agarwal and D. Belgrave and K. Cho and A. Oh},
 pages = {24824--24837},
 publisher = {Curran Associates, Inc.},
 title = {{Chain-of-Thought Prompting Elicits Reasoning in Large Language Models}},
 volume = {35},
 year = {2022}
}

@inproceedings{Zhang:BERTScore:ICLR:2020,
    title={{BERTScore: Evaluating Text Generation with BERT}},
    author={Tianyi Zhang* and Varsha Kishore* and Felix Wu* and Kilian Q. Weinberger and Yoav Artzi},
    booktitle={International Conference on Learning Representations (ICLR)},
    year={2020},
    url={https://openreview.net/forum?id=SkeHuCVFDr}
}

@misc{Wu:LLM-GR:arXiv:2024,
      title={{Thinking with Knowledge Graphs: Enhancing LLM Reasoning Through Structured Data}}, 
      author={Xue Wu and Kostas Tsioutsiouliklis},
      year={2024},
      eprint={2412.10654},
      archivePrefix={arXiv},
      primaryClass={cs.CL},
}

@INPROCEEDINGS{He:Drain:ICWS:2017,
  author={He, Pinjia and Zhu, Jieming and Zheng, Zibin and Lyu, Michael R.},
  booktitle={2017 IEEE International Conference on Web Services (ICWS)}, 
  title={{Drain: An Online Log Parsing Approach with Fixed Depth Tree}}, 
  year={2017},
  volume={},
  number={},
  pages={33-40},
  doi={10.1109/ICWS.2017.13}}

@inproceedings{Wang:PlanSolve:ACL:2023,
    title = {{Plan-and-Solve Prompting: Improving Zero-Shot Chain-of-Thought Reasoning by Large Language Models}},
    author = "Wang, Lei  and
      Xu, Wanyu  and
      Lan, Yihuai  and
      Hu, Zhiqiang  and
      Lan, Yunshi  and
      Lee, Roy Ka-Wei  and
      Lim, Ee-Peng",
    editor = "Rogers, Anna  and
      Boyd-Graber, Jordan  and
      Okazaki, Naoaki",
    booktitle = "Proceedings of the Association for Computational Linguistics (ACL)",
    month = jul,
    year = "2023",
    volume = 1,
    address = "Toronto, Canada",
    publisher = "Association for Computational Linguistics",
    doi = "10.18653/v1/2023.acl-long.147",
    pages = "2609--2634",
}

@article{Pukelsheim:3sigma:1994,
  title={{The Three Sigma Rule}},
  author={Pukelsheim, Friedrich},
  journal={The American Statistician},
  volume={48},
  number={2},
  pages={88--91},
  year={1994},
  publisher={Taylor \& Francis},
  doi = {10.1080/00031305.1994.10476030}
}

@INPROCEEDINGS{Liu:IsolationForest:2008,
  author={Liu, Fei Tony and Ting, Kai Ming and Zhou, Zhi-Hua},
  booktitle={IEEE International Conference on Data Mining}, 
  title={{Isolation Forest}}, 
  year={2008},
  volume={},
  number={},
  pages={413-422},
  doi={10.1109/ICDM.2008.17}
}

@inproceedings{Golovnenva:Roscoe:ICLR:2023,
    title={{{ROSCOE}: A Suite of Metrics for Scoring Step-by-Step Reasoning}},
    author={Olga Golovneva and Moya Peng Chen and Spencer Poff and Martin Corredor and Luke Zettlemoyer and Maryam Fazel-Zarandi and Asli Celikyilmaz},
    booktitle={The Eleventh International Conference on Learning Representations (ICLR)},
    year={2023},
    url={https://openreview.net/forum?id=xYlJRpzZtsY}
}

@inproceedings{Prasad:ReCEval:EMNLP:2023,
    title = {{{R}e{CE}val: Evaluating Reasoning Chains via Correctness and Informativeness}},
    author = "Prasad, Archiki and Saha, Swarnadeep and Zhou, Xiang and Bansal, Mohit",
    editor = "Bouamor, Houda  and Pino, Juan  and Bali, Kalika",
    booktitle = "Conference on Empirical Methods in Natural Language Processing (EMNLP)",
    month = dec,
    year = "2023",
    address = "Singapore",
    publisher = "Association for Computational Linguistics",
    doi = "10.18653/v1/2023.emnlp-main.622",
    pages = "10066--10086",
}

@inproceedings{Mondorf:ReasoningSurvey:COLM:2024,
    title={{Beyond Accuracy: Evaluating the Reasoning Behavior of Large Language Models - A Survey}},
    author={Philipp Mondorf and Barbara Plank},
    booktitle={{Conference on Language Modeling (COLM)}},
    year={2024},
    url={https://openreview.net/forum?id=Lmjgl2n11u}
}

@misc{Gu:LLMJudgeSurvey:arXiv:2025,
      title={{A Survey on LLM-as-a-Judge}}, 
      author={Jiawei Gu and Xuhui Jiang and Zhichao Shi and Hexiang Tan and Xuehao Zhai and Chengjin Xu and Wei Li and Yinghan Shen and Shengjie Ma and Honghao Liu and Saizhuo Wang and Kun Zhang and Yuanzhuo Wang and Wen Gao and Lionel Ni and Jian Guo},
      year={2025},
      eprint={2411.15594},
      archivePrefix={arXiv},
      primaryClass={cs.CL},
}

@inproceedings{Ye:LLMJudgeBias:ICLR:2025,
    title={{Justice or Prejudice? Quantifying Biases in {LLM}-as-a-Judge}},
    author={Jiayi Ye and Yanbo Wang and Yue Huang and Dongping Chen and Qihui Zhang and Nuno Moniz and Tian Gao and Werner Geyer and Chao Huang and Pin-Yu Chen and Nitesh V Chawla and Xiangliang Zhang},
    booktitle={ International Conference on Learning Representations (ICLR)},
    year={2025},
    url={https://openreview.net/forum?id=3GTtZFiajM}
}

@inproceedings{Zheng:LLMJudge:NeurIPS:2023,
     author = {Zheng, Lianmin and Chiang, Wei-Lin and Sheng, Ying and Zhuang, Siyuan and Wu, Zhanghao and Zhuang, Yonghao and Lin, Zi and Li, Zhuohan and Li, Dacheng and Xing, Eric and Zhang, Hao and Gonzalez, Joseph E and Stoica, Ion},
     booktitle = {Advances in Neural Information Processing Systems (NeurIPS)},
     editor = {A. Oh and T. Naumann and A. Globerson and K. Saenko and M. Hardt and S. Levine},
     pages = {46595--46623},
     publisher = {Curran Associates, Inc.},
     title = {{Judging LLM-as-a-Judge with MT-Bench and Chatbot Arena}},
     volume = {36},
     year = {2023}
}

@article{Wilcoxon:JSTOR:1945,
  title={{Individual comparisons by ranking methods}},
  author={Wilcoxon, Frank},
  journal={Biometrics Bulletin},
  volume={1},
  number={6},
  pages={80--83},
  year={1945},
  publisher={JSTOR},
  doi={10.2307/3001968}
}

@misc{qwen3,
      title={{Qwen3 Technical Report}}, 
      author={An Yang and Anfeng Li and Baosong Yang and Beichen Zhang and Binyuan Hui and Bo Zheng and Bowen Yu and Chang Gao and Chengen Huang and Chenxu Lv and Chujie Zheng and Dayiheng Liu and Fan Zhou and Fei Huang and Feng Hu and Hao Ge and Haoran Wei and Huan Lin and Jialong Tang and Jian Yang and Jianhong Tu and Jianwei Zhang and Jianxin Yang and Jiaxi Yang and Jing Zhou and Jingren Zhou and Junyang Lin and Kai Dang and Keqin Bao and Kexin Yang and Le Yu and Lianghao Deng and Mei Li and Mingfeng Xue and Mingze Li and Pei Zhang and Peng Wang and Qin Zhu and Rui Men and Ruize Gao and Shixuan Liu and Shuang Luo and Tianhao Li and Tianyi Tang and Wenbiao Yin and Xingzhang Ren and Xinyu Wang and Xinyu Zhang and Xuancheng Ren and Yang Fan and Yang Su and Yichang Zhang and Yinger Zhang and Yu Wan and Yuqiong Liu and Zekun Wang and Zeyu Cui and Zhenru Zhang and Zhipeng Zhou and Zihan Qiu},
      year={2025},
      eprint={2505.09388},
      archivePrefix={arXiv},
      primaryClass={cs.CL},
}

@misc{llama33,
  title        = {Llama 3.3},
  author       = {Meta AI},
  year         = {2024},
  howpublished = {\url{https://github.com/meta-llama/llama-models/blob/main/models/llama3_3/MODEL_CARD.md}},
  note         = {Accessed 2025-11-04}
}

@misc{llama32,
  title        = {Llama 3.2},
  author       = {Meta AI},
  year         = {2024},
  howpublished = {\url{https://github.com/meta-llama/llama-models/blob/main/models/llama3_2/MODEL_CARD.md}},
  note         = {Accessed 2025-11-04}
}

@misc{command-r-plus,
  title  = {{Command R+}},
  author = {{Cohere}},
  year = {2024},
  howpublished = {\url{https://docs.cohere.com/v2/docs/command-r-plus}},
  note   = {Accessed 2025-11-04},
}

@misc{deepseekr1,
      title={{DeepSeek-R1: Incentivizing Reasoning Capability in LLMs via Reinforcement Learning}}, 
      author={DeepSeek-AI},
      year={2025},
      eprint={2501.12948},
      archivePrefix={arXiv},
      primaryClass={cs.CL},
}

@misc{gpt5,
	title = {{Introducing {GPT}-5}},
    author= {OpenAI},
	howpublished = {\url{https://openai.com/index/introducing-gpt-5/}},
	year = {2025},
    note   = {Accessed 2025-11-04},
}

@book{Sridharan:Observability:2018,
  title     = {{Distributed Systems Observability}},
  author    = {Cindy Sridharan},
  year      = {2018},
  isbn      = {9781492033424},
  publisher = {O'Reilly Media, Inc.}
}

@book{Corbin:OpenCoding:2025,
	address = {Thousand Oaks, California},
	title = {Basics of {Qualitative} {Research} (3rd ed.): {Techniques} and {Procedures} for {Developing} {Grounded} {Theory}},
	publisher = {SAGE Publications, Inc.},
	author = {Corbin, Juliet and Strauss, Anselm},
	year = {2025},
	doi = {10.4135/9781452230153},
}

@software{langgraph,
  author = {LangChain AI},
  title = {{LangGraph}},
  url = {https://langchain-ai.github.io/langgraph/},
  year = {2024},
  note = {Accessed: 2025-11-05}
}

@mastersthesis{thesis,
    author = {Evelien Riddell},
    title = {{Grounded or Guessing? An Empirical Evaluation of LLM Reasoning in Agentic Workflows for Root Cause Analysis in Cloud-based Systems}},
    school = {{University of Waterloo}},
    year = {2026},
    url = {https://hdl.handle.net/10012/22841}
}

@misc{appendix,
    title = {{Online Repository and Supplementary Material}},
    author={Evelien Riddell},
    year = {2026},
    url = {https://github.com/boerste/rca-llm-reasoning}
}

\newpage
\clearpage
\appendix

\section{Supplementary Material} \label{sec:appendix-data}

In this appendix, we provide supplementary material and additional data to support the experiments.
All code, experimental results, and supporting documents for this work can be found in our online repository~\cite{appendix}.

\subsection{Detailed Problem Definition} \label{sec:problem-def-long}
We focus on (1) root cause localization, (2) fault type classification, and (3) propagation path identification based on multi-modal telemetry data.
Given a set of failures (i.e., multi-modal alert events) caused by a single fault, the agent's objective is to jointly perform the three specified tasks.

We model a heterogeneous cloud-based software system as a typed knowledge graph 
$ \mathcal{G} = (\mathcal{E}, \mathcal{R}, \mathcal{T}_\mathcal{E}, \mathcal{T}_\mathcal{R})$
which encodes explicit domain knowledge and operational dependencies within the system,
where $\mathcal{E}$ 
is the set of system entities composed of entity types $t\in \mathcal{T}_\mathcal{E}$ (e.g., services, hosts, data stores)
and $\mathcal{R} \subseteq \mathcal{E} \times \mathcal{T}_\mathcal{R} \times \mathcal{E}$ 
is the set of typed edges
with types drawn from the relationships in $\mathcal{T}_\mathcal{R}$ 
(e.g., \textit{control-flow}, \textit{hosted-on}, \textit{instance-of}). 
Each entity type $t \in \mathcal{T}_\mathcal{E}$ is associated with a set of fault classes $\mathcal{F}_t$.
%

%
For a given fault $s$ and its associated multi-modal monitoring data (metrics, traces, and logs), we assume a set of alerts $\mathcal{A}_s$, each mapped to an element in the knowledge graph, as determined by an external alert extraction procedure (\Cref{sec:alert-extraction}).
Specifically, each alert $a \in \mathcal{A}_s$ is associated with a graph element, i.e., either an entity or relationship, denoted as $c(a) \in \mathcal{E} \cup \mathcal{R}$.
Given a set of alerts for fault $s$, the LLM agent is prompted to produce a ranked list of $k$ fault hypotheses:
\[
\mathcal{H}_s[k] = \{h_s^{(i)}\} _{i=1} ^ k.
\]
Each hypothesis $h^{(i)}_s$, is a structured tuple
\[
h^{(i)} = (e^{(i)}, f^{(i)}, p^{(i)}, j^{(i)}),
\]
where $e^{(i)}\in \mathcal{E}$ is the predicted root-cause entity, 
$f^{(i)} \in \mathcal{F}_t$ is the predicted fault type for $e^{(i)}$ with $t$ being the type of $e^{(i)}$, $p^{(i)}$ is a path in the knowledge graph, and $j^{(i)}$ is the natural language justification.
Each propagation path $p^{(i)}$, defined as
\[
p^{(i)} = [r_{e^{(i)}, \xi}, r_{\xi, \xi-1}, ..., r_{1, c(a)}]
\]
must correspond to a valid walk in the graph, 
where each step $r_{m,n} = (e_m, \tau_{m,n}, e_n)$ corresponds to an edge $r \in \mathcal{R}$ with type $\tau \in \mathcal{T}_\mathcal{R}$, and must terminate at an alerted entity or relationship $c(a)$.
The agent provides a ranked list of human-interpretable hypotheses, each explaining a likely root cause, its fault type, and its propagation path through the system graph.



\subsection{Dataset Filtering and Balancing}

To ensure that fault records used in our analysis were temporally sound and supported by multi-modal telemetry, we applied a sequence of filtering steps to \A. Starting from 16{,}205 injected fault records, we first removed temporally overlapping injections by discarding any fault whose duration intersected the preceding one or whose inter-fault gap was shorter than 45 seconds, yielding 12{,}522 (77.3\%) non-overlapping faults. We then identified extended telemetry gaps---periods with more than 30 minutes of missing data---and excluded fault records whose duration overlapped these gaps, resulting in 4{,}343 (26.8\%) fault records.

Because the two datasets differ substantially in size and class composition, we next balanced \A~so that its final scale approximated that of \B~and no single fault class dominated. To alleviate sparsity among minority classes, we minimally augmented rare faults by introducing time-shifted duplicates placed in non-overlapping synthetic windows. We then downsampled the majority classes, producing a final \A~dataset that combines original minority-class samples, their augmented variants, and a controlled subset of majority-class faults. This yielded 152 faults in total (0.94\% of the original; 3.5\% of the filtered set), as reported in \Cref{tab:dataset-details}. No additional balancing was required for \B, since the dataset was already balanced by the authors of~\cite{Xu:OpenRCA:ICLR:2025}.
The final per–fault-type distributions for \A~and \B~are summarized in \Cref{tab:dataset-a-faults,tab:dataset-b-faults}.

\begin{table}[t]
\footnotesize
\centering
\caption{Distribution of Injected Faults for Dataset \A}
\label{tab:dataset-a-faults}
    \begin{tabular}{l c c c}
        \toprule
        \textbf{Fault Type} & \textbf{Count} & \textbf{\makecell{Injection\\Locations}} & \textbf{Fault Granularity} \\
        \midrule
        High Memory Usage & 40 (26\%)& 10 & Service inst.\\
        Session Timeout  & 40 (26\%) & 2 & Service inst.\\
        File Missing  & 36 (24\%) & 2 & Service inst.\\
        \makecell[l]{Unexpected Process\\Termination} & 20 (13\%) & 4 & Service inst.\\
        \makecell[l]{Internal Permission\\Misconfiguration} & 16 (11\%) & 2 & Service inst.\\
        \bottomrule
    \end{tabular}
\end{table}
\begin{table}[t]
    \centering
    \footnotesize
    \setlength{\tabcolsep}{3.5pt}
    \caption{Distribution of Injected Faults for Dataset \B}
    \label{tab:dataset-b-faults}
    \begin{tabular}{l c c c}
        \toprule
        \textbf{Fault Type} & \textbf{Count} & \textbf{\makecell{Injection\\Locations}} & \textbf{Fault Granularity} \\
        \midrule
        Container CPU Load & 13 (9\%) & 12 & Service, service inst. \\
        Container Memory Load & 13 (9\%) & 11 & Service, service inst. \\
        Container Packet Corruption & 13 (9\%)& 12 & Service, service inst. \\
        \makecell[l]{Container Packet Retransmission} & 13 (9\%) & 12 & Service, service inst. \\
        Container Packet Loss & 8 (5\%) & 7 & Service, service inst. \\
        Container Network Latency & 8 (5\%) & 8 & Service, service inst. \\
        Container Process Termination & 7 (5\%) & 6 & Service, service inst. \\
        Container Read I/O Load & 17 (11\%) & 12 & Service, service inst. \\
        Container Write I/O Load & 5 (3\%) & 5 & Service, service inst. \\
        Node Disk Write I/O & 10 (7\%) & 3 & Host \\
        Node Disk Read I/O & 9 (6\%) & 6 & Host \\
        Node Disk Space Consumption & 10 (7\%) & 6 & Host \\
        Node Memory Consumption & 9 (6\%) & 5 & Host \\
        Node CPU Load & 9 (6\%) & 4 & Host \\
        Node CPU Spike & 4 (3\%) & 3 & Host \\
        \bottomrule
    \end{tabular}
\end{table}

\subsection{System Knowledge Graph}

We represent the system as the typed knowledge graph 
\[ \mathcal{G}=(\mathcal{E},\mathcal{R},\mathcal{T}_\mathcal{E},\mathcal{T}_\mathcal{R}). \]
Nodes $e \in \mathcal{E}$ (with types in $\mathcal{T}_\mathcal{E}$) represent system entities.
Typed edges $r \in \mathcal{R}$ (with types in $\mathcal{T}_\mathcal{R}$) capture dependencies and interactions commonly observed in cloud-native architectures, such as control and data flow dependencies, instantiations of software components to capture redundancy and scaling
(e.g., \textit{instance-of}), and deployment relations (e.g., \textit{hosted-on}).
This formalization allows us to (i) associate each entity type $t$ with its fault class set \(\mathcal{F}_t\)\footnote{
Some entity types in some datasets have an empty $\mathcal{F}_t$. For example, Dataset \A~contains faults only for \textit{Service-Instance}-type entities, while the system is comprised of additional types of entities.
}
and (ii) ground alerts $a \in \mathcal{A}_s$ to graph elements via the mapping \(c(a)\in\mathcal{E}\cup\mathcal{R}\).

\begin{figure}[tb]
    \centering
    \includegraphics[width=0.75\linewidth]{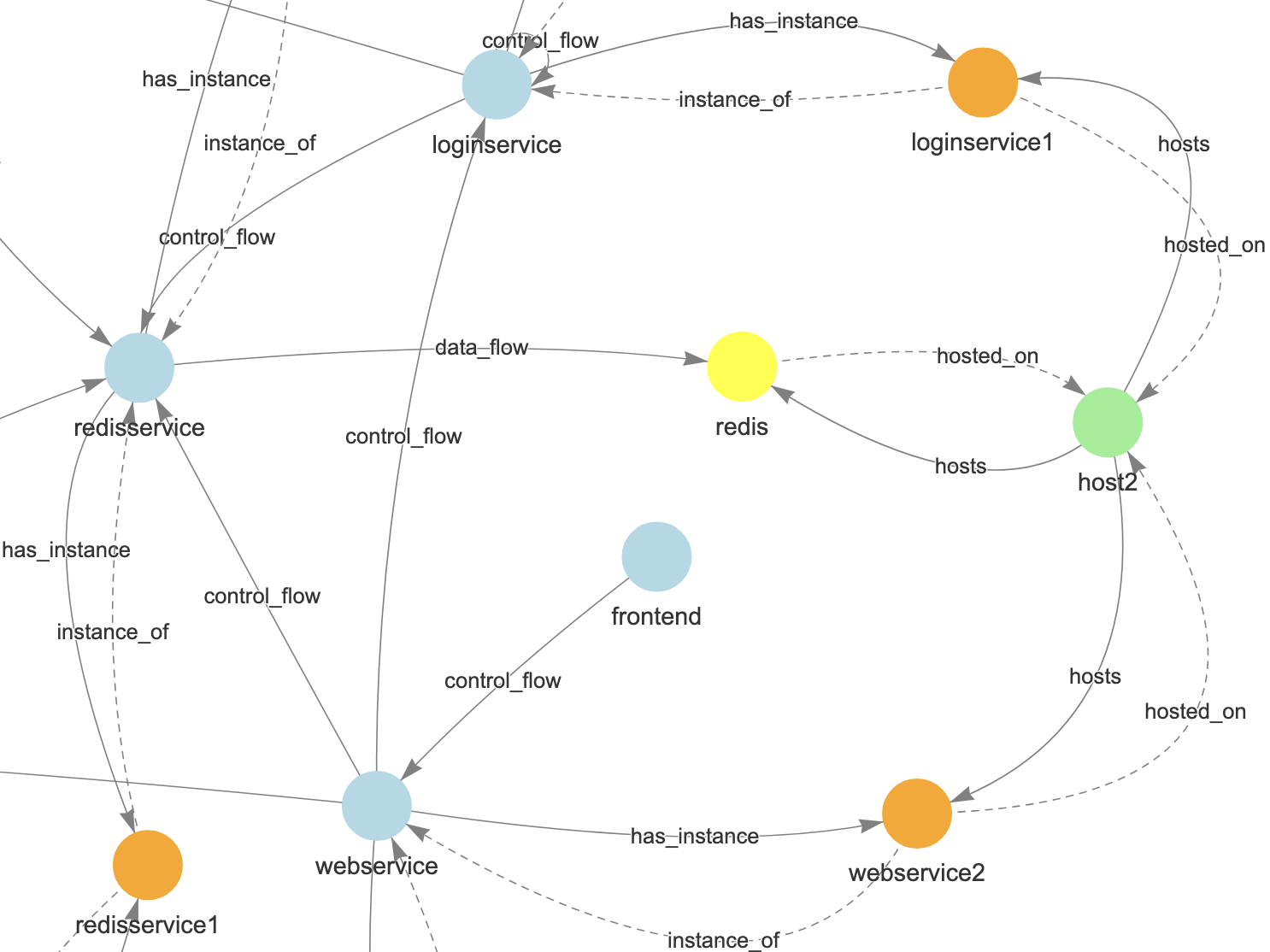} 
    \caption{A region of the system KG for MicroSS~\cite{GAIA:dataset}. 
    Blue, orange, green, and yellow nodes represent \textit{Service}, \textit{Service Instance}, \textit{Host}, and \textit{Cache} type entities, respectively. Only \textit{instance-of}, \textit{has-instance}, \textit{hosted-on}, \textit{hosts}, \textit{control-flow}, and \textit{data-flow} relationships are shown for simplicity.
    }
    \label{fig:MicroSS-KG}
\end{figure}

\Cref{tab:system-entities} lists the entity specifications ($\mathcal{T}_\mathcal{E}$) used for our benchmark datasets.
Additionally,
\Cref{fig:MicroSS-KG} provides an example of a region of the KG constructed for MicroSS.
In this case, entity types include \textit{Service, Service-Instance, Orchestration-Manager, Host, Database} and \textit{Cache}.
For simplicity, we show a subset of relationships.

\subsection{Agent Tools}

We summarize the agent tools described in~\Cref{sec:tools} in more detail in~\Cref{tab:tools}.

\subsection{Reasoning Failure Taxonomy} \label{sec:appendix-rf-taxonomy}
\begin{figure*}
    \centering
    \includegraphics[width=\textwidth]{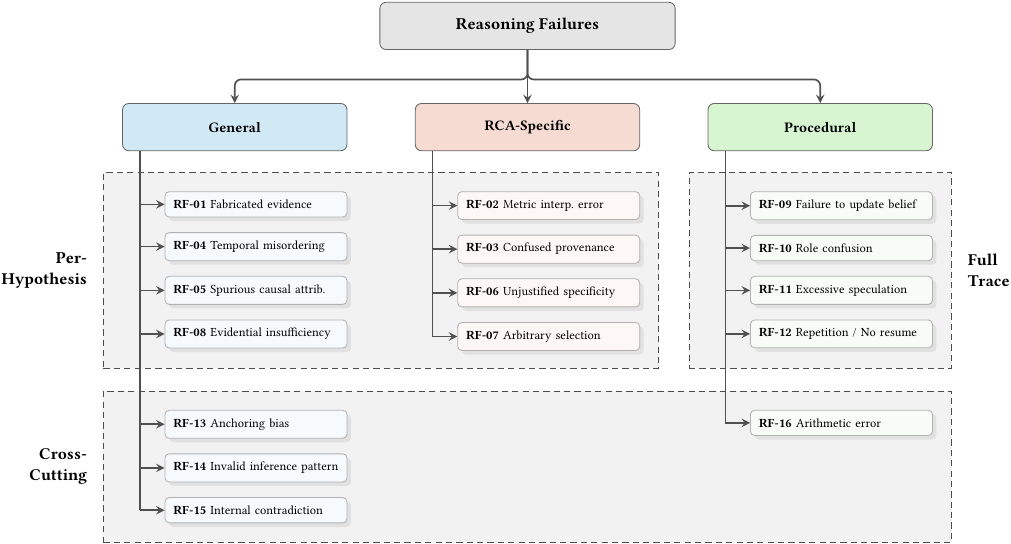}
    \vspace{1pt}
    \caption{Reasoning failures taxonomy.
    Columns partition failure \emph{categories} and horizontal swimlanes indicate the \emph{scope} in which the failure manifests.}
    \label{fig:rf-swimlane-diagram}
    \vspace{8pt}
\end{figure*}

\Cref{fig:rf-swimlane-diagram} provides a visual representation of the reasoning failure taxonomy.
Columns partition failure \emph{categories} and horizontal swimlanes indicate the \emph{scope} in which the failure manifests: per-hypothesis (local errors during evaluation of a single hypothesis), full-trace (errors that emerge across the agent's whole reasoning trajectory), and cross-cutting (systemic biases or errors that pervade multiple stages). 
Reasoning failure (RF) codes correspond to those introduced in \Cref{tab:reasoning-failures-taxonomy}.

\subsection{Results}

\subsubsection{Accuracy}
\begin{figure*}[!h]
    \centering
    \begin{minipage}{0.49\textwidth}
        \centering
        \includegraphics[width=\textwidth]{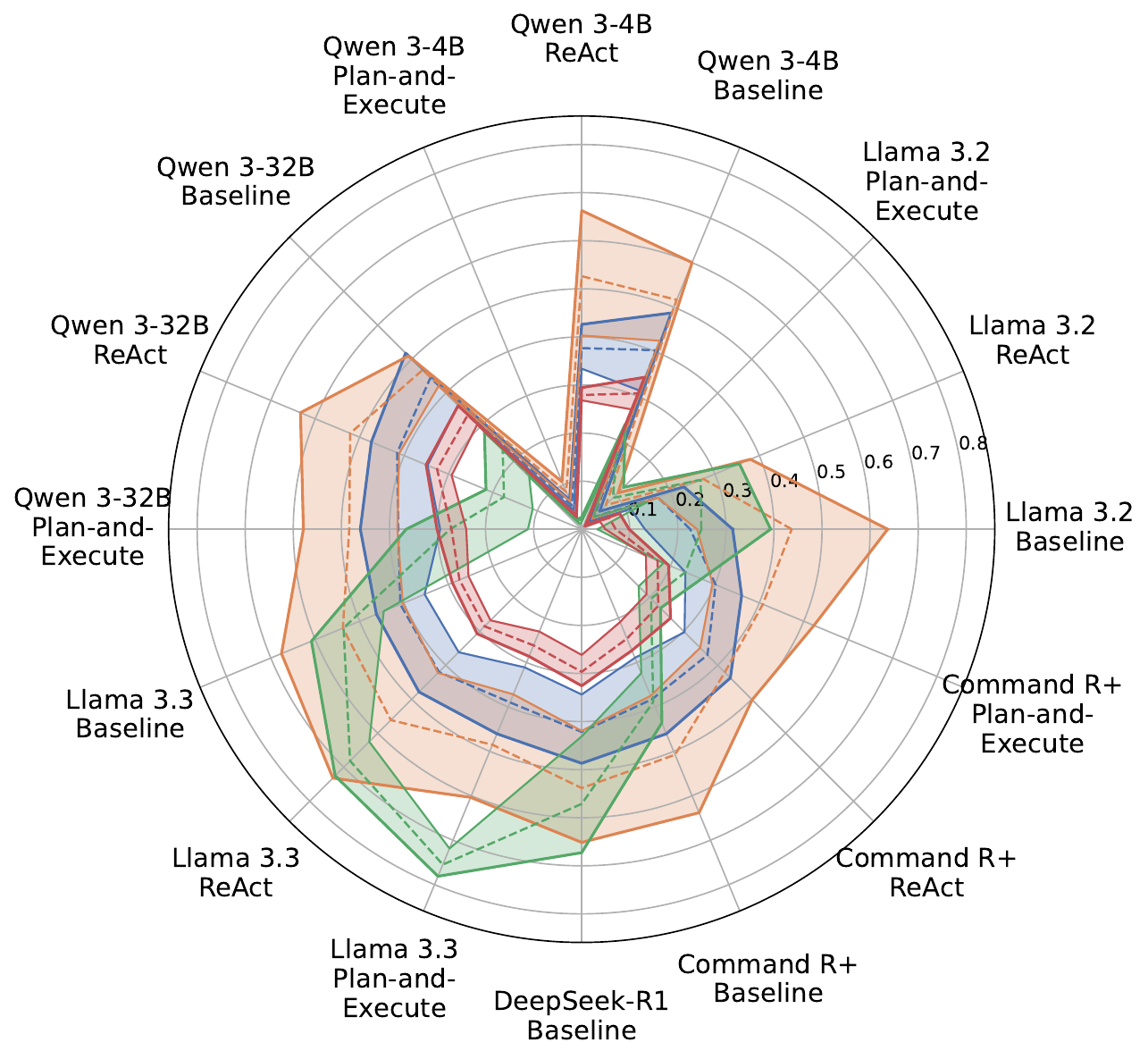}
        \vspace{-2pt}
        \small (a) \A
        \label{fig:accuracy-radar-A}
    \end{minipage}
    \begin{minipage}{0.5\textwidth}
        \centering
        \includegraphics[width=\textwidth]{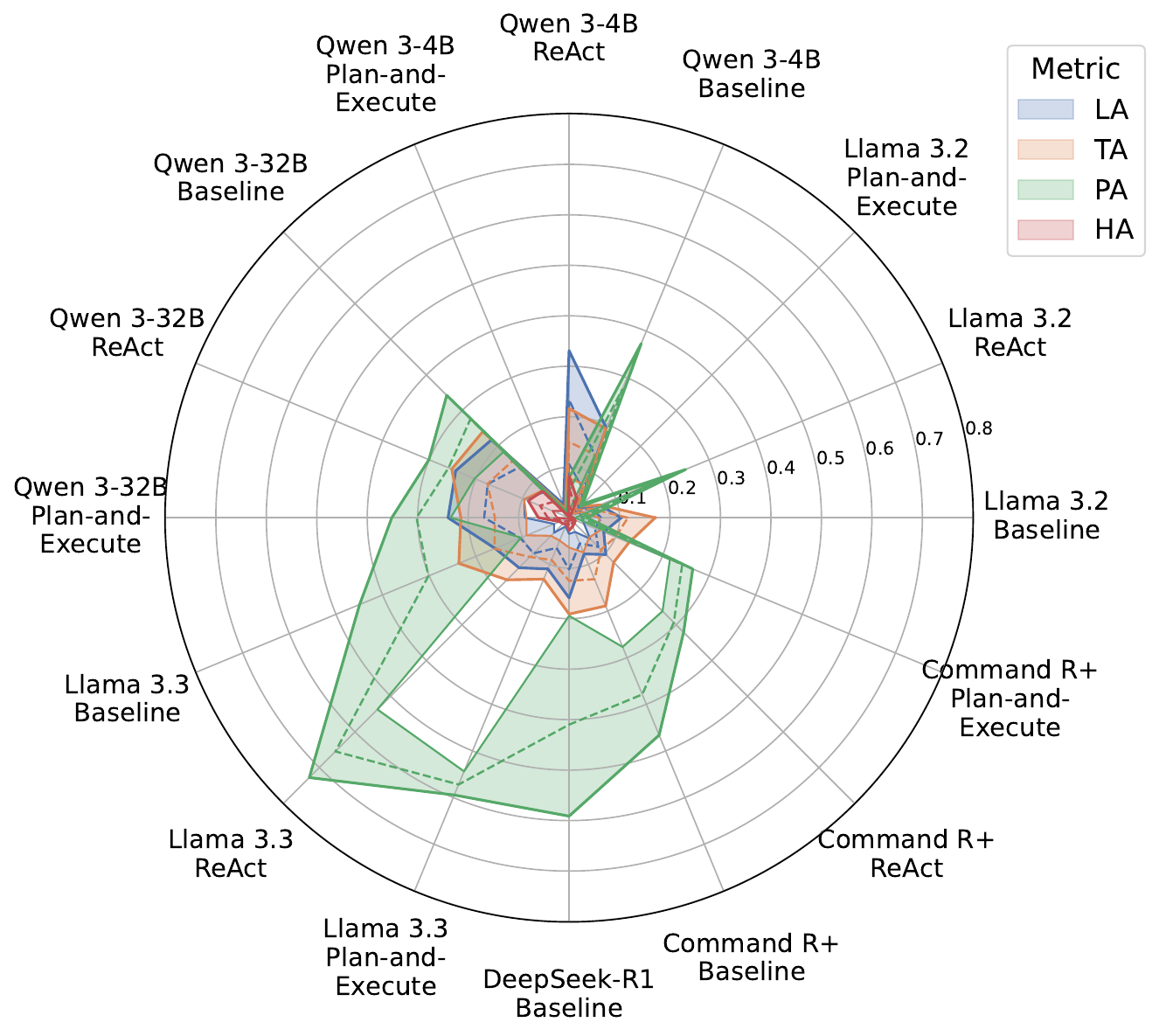}
        \vspace{-2pt}
        \small (b) \B
        \label{fig:accuracy-radar-B}
    \end{minipage}
    \vspace{5pt}
    \caption{Accuracy results for datasets (a) \A~and (b) \B~(lower bound is A@1, upper bound is A@3, and dotted line is Avg@3).}
    \label{fig:accuracy-radar-datasets}
\end{figure*}

To complement the results reported in \Cref{tab:accuracy-results} (\Cref{sec:results-rq1}), 
\Cref{fig:accuracy-radar-datasets} presents per-dataset accuracy plots.

\begin{table}[!t]
\centering
\footnotesize
\caption{Description of Tools}
\label{tab:tools}
\begin{tabularx}{\linewidth}{@{}clX@{}}
\toprule
\textbf{Category} & \multicolumn{1}{c}{\textbf{Tool}} & \multicolumn{1}{c}{\textbf{Description}} 
\\ \midrule
\multirow{4}{*}{\makecell{Data\\Characteristics}} 
  & \text{check\_node\_existence} & Check if a named entity exists in the system. \\
  & \text{get\_node\_attributes} & Retrieve attributes and alert data of a given entity. \\
  & \makecell[l]{\text{get\_all\_instances\_of\_}\\\text{entity\_type}} & Enumerate all instances of a specified entity type. \\
  & \text{get\_edge\_attributes} & Inspect properties of edges between two entities. \\ \midrule
\multirow{2}{*}{\makecell{Graph\\Traversal}} 
  & \text{get\_node\_neighborhood} & Retrieve the r-hop neighborhood of a given entity. \\
  & \text{get\_all\_simple\_paths} & Enumerate all simple paths between two entities. \\
\bottomrule
\end{tabularx}
\end{table}
\begin{table}[!t]
\centering
\footnotesize
\caption{Average LLM Inference Time Per Scenario}
\label{tab:ttr}
\begin{tabular}{lccc}
\toprule
\textbf{Model} 
& \makecell{\textbf{\baseline} \\ \textit{min}} 
& \makecell{\textbf{\react} \\ \textit{min (\%$\Delta$)}} 
& \makecell{\textbf{\pe} \\ \textit{min (\%$\Delta$})} \\
\midrule
Llama 3.2 (3B) & 0.76 & 1.8 (+139.1\%) & 8.3 (+987.0\%) \\
Qwen 3 (4B)    & 3.2 & 4.8 (+49.1\%)  & 6.1 (+88.2\%) \\
Qwen 3 (32B)   & 5.0 & 5.0 (+0.9\%)   & 10.8 (+116.3\%) \\
Llama 3.3      & 6.0 & 5.7 (--4.9\%)  & 10.7 (+78.3\%) \\
DeepSeek-R1 & 4.3 & N/A & N/A \\
Command R+     & 4.6 & 5.4 (+17.6\%)  & 27.3 (+499.5\%) \\
\bottomrule
\end{tabular}
\par\smallskip
\begin{minipage}{0.88\linewidth}
    \footnotesize
    Time in minutes (\textit{min}); parentheses show percent change ($\%\Delta$) relative to \baseline.
\end{minipage}
\end{table}

\subsubsection{LLM Inference Times} \label{sec:appendix-llm-ttr}

To provide context on the operational cost of using different workflows, we analyze the inference time (i.e., time-to-result) for all evaluated models. \Cref{tab:ttr} details the average inference time per scenario in minutes across workflows. The percent change (\%$\Delta$) is calculated relative to the \baseline~baseline.
Our experiments show that \react~and \pe~increased average inference time by 40\% and 354\%, respectively, compared to the \baseline~baseline.
\begin{table*}[th]
  \centering
  \footnotesize
  \caption{High-level System Entity Types}  
  \renewcommand{\arraystretch}{1.3}
  \label{tab:system-entities}
  \begin{tabularx}{0.9\textwidth}{p{1.2cm} p{2.3cm} X p{3cm}}
    \toprule
    \textbf{Category} & \textbf{Entity Types} & \textbf{Description} & \textbf{Examples} \\
    \midrule
    \multirow[c]{5}{*}{Software}
        & Service & An aggregation of software that satisfies an end-use function. & (Micro)service, API \\
        & Database & An organized persistent collection of data and information that allows for its retrieval. & MySQL, key-value store, file \\
        & Cache & An organized non-persistent collection of data and information that allows for its retrieval. & Redis \\
        & Coordination Manager & Manages metadata, state synchronization, and coordination tasks. & ZooKeeper, Consul \\
        & Software Component Instance & A specific and identifiable runtime execution of the \textit{Software Component} (i.e., service, database, cache). & Service instance, database instance, cache instance\\
    \midrule
    \multirow[c]{1}{*}{Hardware} 
        & Host & A virtual or physical computer, hardware device, or environment where software components or programs are deployed, installed, or executed. & Virtual machine, container, server \\
    \bottomrule
  \end{tabularx}
  \vspace{6pt}
\end{table*}

\subsubsection{Sensitivity to Input Representation Strategies} \label{sec:appendix-input-rep}

\begin{figure}[!tbh]
    \centering
    \begin{subfigure}{\linewidth}
        \centering
        \includegraphics[width=\linewidth]{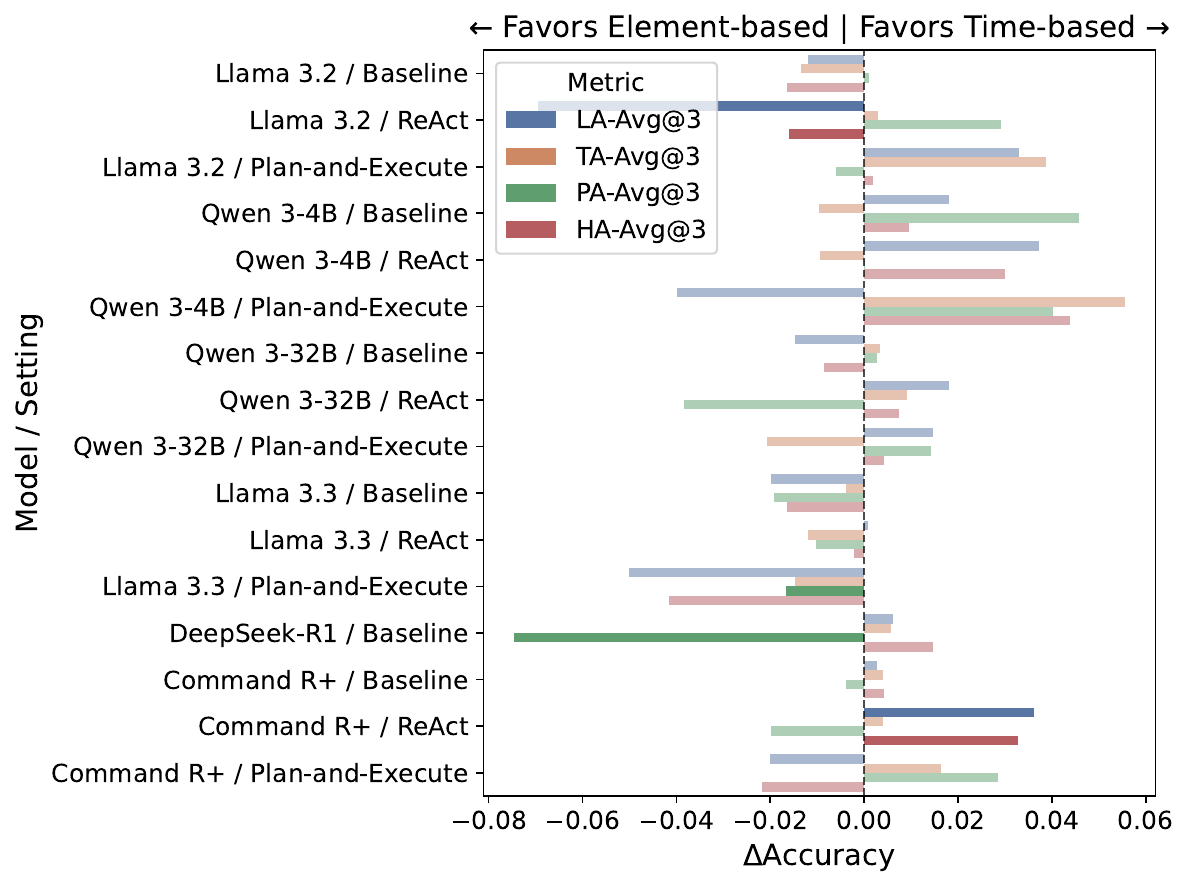}
        \caption{\textit{Alert unification} strategies}
        \label{fig:accuracy-alert-representation}
    \end{subfigure}
    \vspace{3pt}
    \begin{subfigure}{\linewidth}
        \centering
        \includegraphics[width=\linewidth]{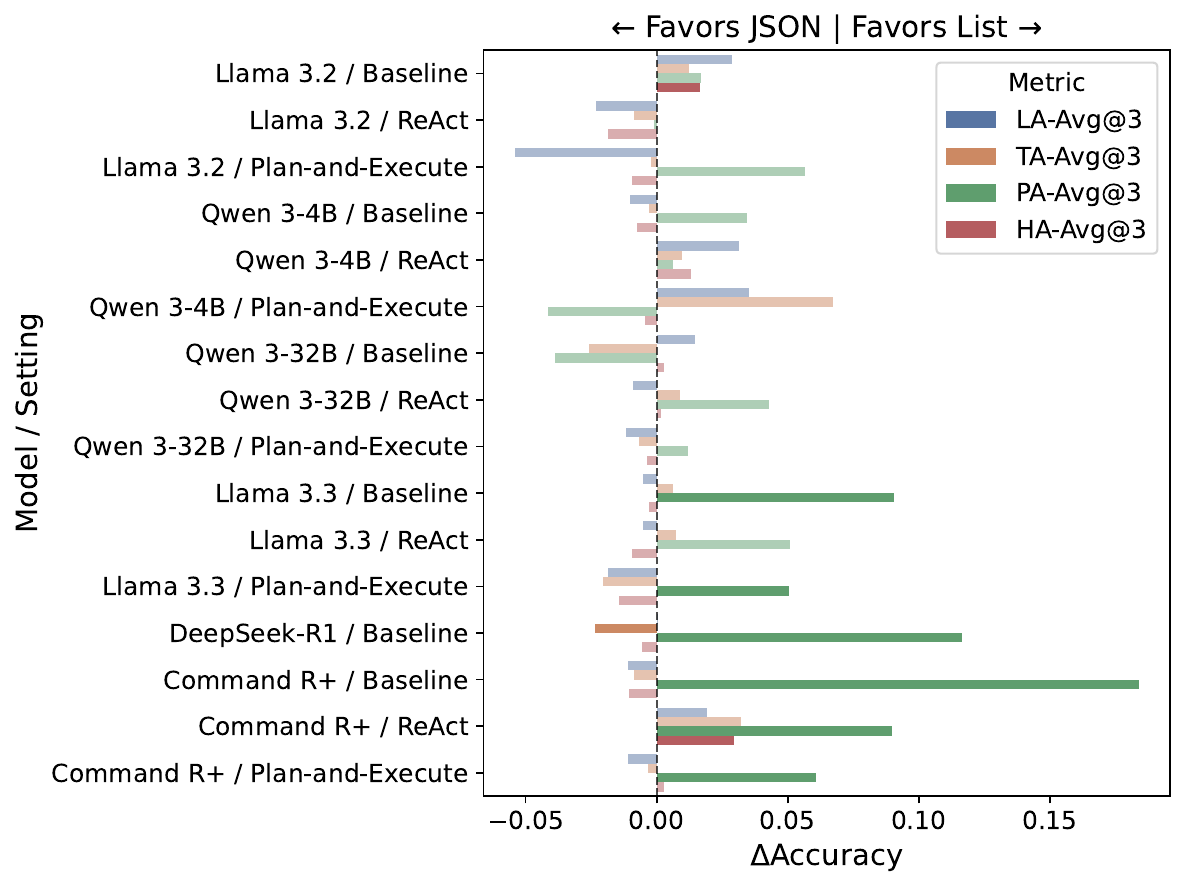}
        \caption{\textit{KG representation} strategies}
        \label{fig:accuracy-kg-representation}
    \end{subfigure}
    \caption{Change in accuracy ($\Delta \text{A-Avg}@3$) for (a) \textit{alert unification} and (b) \textit{KG representation} strategies. 
    In both figures, \textbf{solid bars} denote statistically significant differences ($p < 0.05$) according to the Wilcoxon signed-rank test, while \textbf{faded bars} indicate changes that are not statistically significant.}
    \label{fig:accuracy-input-representations}
    \vspace{3pt}
\end{figure}

In our experiments,
we consider two orthogonal aspects of input construction: (1) the \textit{alert unification strategy} and (2) the \textit{KG representation strategy}.
For (1), we compare \textit{time-based} unification, which preserves the chronological sequence of alerts, and \textit{element-based} unification, which groups alerts by their corresponding KG element.
For (2), we compare a \textit{list}-based representation with a \textit{JSON}-based representation of nodes and edges.
Accuracy differences between strategies are reported as $\Delta\text{A-Avg}@3$ in \Cref{fig:accuracy-input-representations}. Statistical significance is assessed using the Wilcoxon signed-rank test~\cite{Wilcoxon:JSTOR:1945}. In both figures, solid bars denote changes significant at $p < 0.05$, while faded bars denote non-significant differences.

\Cref{fig:accuracy-alert-representation} shows that the optimal alert unification strategy varies substantially by model and workflow, and that significant effects are relatively uncommon.
Under \react, Llama~3.2 displays a statistically significant preference for \textit{element-based} unification (negative~$\Delta$).
Conversely, Command~R+ favours time-based unification (positive~$\Delta$), suggesting that it benefits more from preserving temporal proximity when inferring causality.
Notably, DeepSeek-R1 exhibits a preference for element-based grouping for PA ($\Delta$PA $\approx -0.07$), indicating that grouping alerts by system topology aids in reconstructing fault-propagation paths.
Overall, the mixed trends suggest that the efficacy of a unification strategy reflects underlying training biases---temporal versus categorical structuring---rather than a universally optimal format.

The results in \Cref{fig:accuracy-kg-representation} reveal a strong, statistically significant preference for \textit{list-based} KG representations for Llama~3.3, DeepSeek-R1, and Command~R+ for inferring failure-propagation paths.
These models exhibit large positive gains for PA (e.g., Command~R+ $\Delta$PA $> 0.15$), indicating that the concise, token-efficient structure of adjacency lists is substantially more suitable for traversal-style reasoning than the more verbose JSON format.
However, this advantage in pathfinding does not consistently translate to other metrics, where results are mixed or negligible.
Taken together, the findings suggest that while list-based KGs substantially aid models in ``walking'' the graph, they do not consistently improve the model's ability to recover non-topological attributes of the root-cause fault.

\subsection{Prompts} \label{sec:appendix-prompts}

\Cref{fig:input-prompt} shows the RCA input-prompt template we used for the \react\ workflow. 
\Cref{fig:judge-prompt} contains the full LLM-as-a-Judge prompt template, including the failure taxonomy and the step-by-step annotation guidelines provided to the judge model.

\begin{figure}[bh]
\centering
\label{fig:input-prompt}
\begin{tcolorbox}[
    colback=gray!5,       
    colframe=black,       
    width=\linewidth,     
    sharp corners,        
    boxrule=0.5pt,        
    fonttitle=\bfseries,  
    fontupper=\scriptsize,
    left=5pt,
    right=5pt,
    top=5pt,
    bottom=5pt
]
You are a helpful assistant that is an expert in root cause analysis for complex cloud-based software systems.\\

Consider a cloud-based software system composed of multiple interconnected components (both software and hardware). This system can be represented by an explicit, directed, unweighted, and typed knowledge graph, where nodes represent system components and edges indicate relationships between them. The schema of the knowledge graph is as follows.\\

\#\#\# Knowledge graph schema\\
\#\#\#\# Entities\\
\{entity schema\}\\
\#\#\#\# Relationships\\
\{relationship schema\}\\


Errors or issues originating in one component may propagate to others due to dependencies, communication links, or shared resources. These errors often manifest as observable symptoms (e.g., anomalies or alerts) in different system components. \\

\#\#\# Task \\
You will be given a set of symptoms (e.g., log, trace and/or metric alerts) due to a fault that occurred in the system. \\
Your task is to use the system knowledge graph and the detected symptoms to hypothesize the three most likely root cause faults that could be the underlying cause of the observed symptoms. \\
Each root cause fault is localized to a single system component, included as a node in the knowledge graph. You must identify this node as the root cause location. \\
    

\#\#\# Instructions: \\
You should think step-by-step in order to fulfill the objective. The step-by-step workflow should follow a "Thought/Action/Observation" loop that can repeat multiple times if needed. Here is how you should go about it:\\
1. Thought: reflect internally on the current task, the available information, and what to do next.\\
2. Action: if further information is needed, choose one appropriate tool to call. Any and all ``Thoughts" must be included in the `reasoning' field in the tool input.\\
3. Observation: The tool will return a result, which will be provided to you. \\
Repeat this loop as needed until you have enough information to answer the original task.\\
When ready, output your final answer starting with the prefix 'Final Answer:'. 

Your `Final Answer' should consist of three likely root cause faults.\\
For each root cause fault, provide:\\
- **Type**: the type of root cause fault. Please restrict your analysis to the following types of root cause faults: \{root cause fault types\}.\\
- **Description**: an explanation of what the root cause fault looks like in the system.\\
- **Location**: the single exact node at which the root cause fault occurs. This node should have the following entity type: \{root cause fault entity types\}.\\
- **Justification**: a step-by-step reasoning based on the given alerts and information from the knowledge graph that explains how the symptoms could occur due to the root cause.\\
- **Propagation path**: the specific propagation path in the knowledge graph that would make the root cause possible, formatted as `node1 --(edge\_label1)--\textgreater node2 --(edge\_label2)--\textgreater node3`.\\

You should rank the three root cause faults in order of most likely to least likely.


\#\#\# Observed symptoms\\
The following symptoms/alerts were detected by an anomaly detector:\\
\{alerts\}\\
Think step by step and ensure your reasoning is traceable through the knowledge graph.


\end{tcolorbox}
\caption{Input prompt template for \react~workflow.}
\end{figure}
 
\clearpage
\begin{tcolorbox}[
    enhanced,
    colback=gray!5,       
    colframe=black,       
    width=\linewidth,     
    boxrule=0.5pt,        
    title=System Message, 
    fonttitle=\bfseries,  
    fontupper=\scriptsize,
    left=5pt,
    right=5pt,
    top=5pt,
    bottom=5pt
]
You are a rigorous assistant with excellent critical thinking skills. Your task is to qualitatively analyze the reasoning of an LLM agent in a root cause analysis task and identify any reasoning failures according to a given taxonomy. Work through the annotation workflow step by step. Only mark failures you are reasonably confident in.
\end{tcolorbox}
\vspace{-20pt}
\begin{tcolorbox}[
    enhanced,
    breakable,
    colback=gray!5,       
    colframe=black,       
    width=\linewidth,     
    boxrule=0.5pt,        
    title=Human Message,  
    fonttitle=\bfseries,  
    fontupper=\scriptsize,
    left=5pt,
    right=5pt,
    top=5pt,
    bottom=5pt
]
\textbf{\#\# REASONING FAILURE TAXONOMY} \\ 

\textbf{\#\#\# RF-01 — Fabricated evidence (hallucinated alerts/metrics/logs/traces)} \\ 

\textbf{Scope}: per-hypothesis (evidence existence)\\ 
\textbf{Definition}: Model asserts the existence of a specific alert/metric/log/trace that cannot be found in the provided alerts/metrics/logs/traces after up to 3 quick scans.\\ 
\textbf{Example}: Claims "\texttt{2025-09-01 12:05 | METRIC | dbservice1 | disk\_io | up}" or "The disk\_io metric was up for dbservice", but no such record (or any reasonably equivalent entry for that alert) exists in the provided alerts.\\ 
\textbf{Signals}: Model quotes an alert absent in the alert set; model uses confident language about a concrete alert that cannot be located.\\ 
\textbf{Annotation rule}: Perform up to 3 quick scans (exact or close fuzzy match on component + metric/endpoint + time) for each piece of evidence mentioned. If no match found, mark RF-01 and paste model claim + NO MATCH FOUND.\\ 
\textbf{Severity}: 1-5\\ 
  - 1 = single fabricated alert/metric/log/trace in 1 hypothesis\\ 
  - 2 = multiple occurrences in 1 hypothesis\\ 
  - 3 = single occurrence in 2 hypotheses\\ 
  - 4 = multiple occurrences in 2 hypotheses\\ 
  - 5 = present in all 3+ hypotheses\\ 

\textbf{\#\#\# RF-02 — Metric-interpretation error (directionality / semantic misread)} \\

\textbf{Scope}: per-hypothesis (evidence interpretation)\\ 
\textbf{Definition}: Misunderstands/misinterprets metric semantics (up $\implies$ $+3\sigma$, down $\implies$ $-3\sigma$), confuses counters/gauges, or inverts meaning.\\ 
\textbf{Example}: "docker\_memory\_rss\_pct is down, indicating high memory usage." (direction inverted: down for this metric means memory measure decreased); "mem\_usage is down, indicating a memory leak" (misinterpretation: a memory leak would typically correlate with increased memory usage).\\ 
\textbf{Signals}: Interpretation directly contradicts the standard metric meaning or contradicts $\pm3\sigma$ (i.e., up/down) semantics for the metric.\\ 
\textbf{Annotation rule}: If the alert exists but interpretation contradicts metric semantics → label RF-02 and paste the model claim(s) + alert(s) it referenced.\\ 
\textbf{Severity}: 1-5\\ 
  - 1 = single metric misinterpretation in 1 hypothesis\\ 
  - 2 = multiple occurrences in 1 hypothesis\\ 
  - 3 = single occurrence in 2 hypotheses\\ 
  - 4 = multiple occurrences in 2 hypotheses\\ 
  - 5 = present in all 3+ hypotheses\\ 

\textbf{\#\#\# RF-03 — Confused provenance (symptom-observer blamed as cause)} \\

\textbf{Scope}: per-hypothesis (provenance)\\ 
\textbf{Definition}: Model treats the component that observed/logged a symptom as the origin/root cause rather than tracing upstream/downstream sources.\\ 
\textbf{Example}: Webservice log contains "an error occurred in a downstream service"; model concludes "webservice is root cause" instead of investigating downstream services.\\ 
\textbf{Signals}: Log text contains explicit downstream/propagation language; model names observer as cause.\\ 
\textbf{Annotation rule}: If evidence indicates an observed downstream symptom and model blames the observer, mark RF-03.\\ 
\textbf{Severity}: 1-5\\ 
  - 1 = single confused provenance instance in 1 hypothesis\\ 
  - 2 = multiple occurrences in 1 hypothesis\\ 
  - 3 = single occurrence in 2 hypotheses\\ 
  - 4 = multiple occurrences in 2 hypotheses\\ 
  - 5 = present in all 3+ hypotheses\\ 

\textbf{\#\#\# RF-04 — Temporal misordering (timeline error)} \\

\textbf{Scope}: per-hypothesis (timestamps)\\ 
\textbf{Definition}: Model assigns causation to an event occurring after the observed effect or otherwise violates alert timestamp ordering.\\ 
\textbf{Example}: A log says background save (BGSAVE) started at 12:20, but multiple memory/I/O anomalies began at 12:15; model claims BGSAVE causedthe earlier anomalies.\\ 
\textbf{Signals}: Claimed cause timestamp is later than the earliest effect timestamp.\\ 
\textbf{Annotation rule}: Extract model-cited timestamps and compare to alerts; if the causal claim violates the real timeline, mark RF-04. If ordering is implied and contradicts alert order, still mark RF-04.\\ 
\textbf{Severity}: 1-5\\ 
  - 1 = single temporal misordering in 1 hypothesis\\ 
  - 2 = multiple occurrences in 1 hypothesis\\ 
  - 3 = single occurrence in 2 hypotheses\\ 
  - 4 = multiple occurrences in 2 hypotheses\\ 
  - 5 = present in all 3+ hypotheses\\ 

\textbf{\#\#\# RF-05 — Spurious causal attribution (weak/unsupported causation OR mechanism depends on nonexistent KG link)} \\

\textbf{Scope}: per-hypothesis (mechanism/causal chain)\\ 
\textbf{Definition}: Model asserts X → Y causation without adequate support, or uses a causal mechanism that depends on knowledge-graph relationships that do not exist. Plausible speculation consistent with the KG and alerts is acceptable and should not be penalized. KG relationships that are close-enough without detracting from the point being made OR aligned more closely with trace alerts should also not be penalized 
(e.g., "\texttt{webservice1 
--(instance\_of)--> webservice 
--(control\_flow)--> redisservice 
--(has\_instance)--> redisservice2}"
vs "\texttt{webservice1 
-(control\_flow)--> redisservice2}")\\ 
\textbf{Example}: Claims node-6 disk writes cause shippingservice-0 latency because \texttt{node-6 --(hosts)--> shippingservice-0}, but that host relationship is nonexistent in the KG.\\ 
\textbf{Signals}: Use of causal language ("caused", "because of", "therefore") plus no plausible KG or alert support, or explicit citation of nonexistent KG edges (based on the information available).\\ 
\textbf{Annotation rule}: Check KG and alerts for the mechanism, if mechanism unsupported or relies on absent KG links, mark RF-05. \\
\textbf{Severity}: 1-5\\ 
  - 1 = single spurious causal attribution in 1 hypothesis\\ 
  - 2 = multiple occurrences in 1 hypothesis\\ 
  - 3 = single occurrence in 2 hypotheses\\ 
  - 4 = multiple occurrences in 2 hypotheses\\ 
  - 5 = present in all 3+ hypotheses\\ 

\textbf{\#\#\# RF-06 — Unjustified instance/granularity specificity} \\

\textbf{Scope}: per-hypothesis (granularity)\\ 
\textbf{Definition}: Model asserts an instance-level root-cause location when evidence supports only service-/node-level effect, unless unique per-instance multi-modal evidence exists.\\ 
\textbf{Example}: Service-wide alerts, but model blames loginservice2 instance without unique evidence.\\ 
\textbf{Signals}: No unique instance differentiator (multi-modal alerts, unique timestamps, volume, frequency).\\ 
\textbf{Annotation rule}: If instance claim lacks per-instance unique evidence, mark RF-06. Only relevant for cases where the root-cause location can be more than only instance-level.\\ 
\textbf{Severity}: 1-5\\ 
  - 1 = present for 1 hypothesis\\ 
  - 3 = present for 2 hypotheses\\ 
  - 5 = present in all 3 hypotheses\\ 

\textbf{\#\#\# RF-07 — Arbitrary / non-systematic evidence selection (bad triage)} \\

\textbf{Scope}: per-hypothesis (evidence selection)\\ 
\textbf{Definition}: Model focuses on a seemingly arbitrary alert subset inconsistent with simple triage heuristics (i.e., first-seen,highest-frequency, highest-volume, multi-modal alerts vs single-modal alerts) without rationale, and selection plausibly alters diagnostic trajectory/conclusions.\\ 
\textbf{Example}: Investigates loginservice2 though loginservice1 has identical metric/trace alerts earlier in time; or investigates mobservice1 while dbservice2 has more metric alerts.\\ 
\textbf{Signals}: Chosen evidence is seemingly arbitrary and does not follow logical selection procedures: earliest/most frequent/highest volume multi-modal.\\ 
\textbf{Annotation rule}: If model's chosen evidence subset is plausibly arbitrary or contradicts simple triage heuristics and that choice affected the top hypothesis or multiple hypotheses → mark RF-07 (higher severity). If it did not materially change outcome → mark with low severity.\\ 
\textbf{Severity}: 1-5\\ 
  - 1 = present for 1 hypothesis\\ 
  - 3 = present for 2 hypotheses\\ 
  - 5 = present in all 3 hypotheses\\ 

\textbf{\#\#\# RF-08 — Evidential insufficiency (supported but weak / non-specific for the claim)} \\

\textbf{Scope}: per-hypothesis (evidence sufficiency)\\ 
\textbf{Definition}: Evidence exists but its temporal precision, frequency, mechanism link, discriminability, provenance clarity, or granularity is insufficient to support the specific claim. This is an inferential-sufficiency error, not a hallucination.\\ 
\textbf{Signals}: Checklist failures: temporal precedence; frequency; mechanism; discriminability; provenance clarity; granularity alignment.\\ 
\textbf{Annotation rule}: After prior per-hypothesis checks, apply Sufficiency Checklist; if required items fail for claim type, mark RF-08 and provide model claim(s) + matched alert(s) + list which checklist items failed.\\ 
\textbf{Severity}: 1-5\\ 
  - 1 = single insufficiency in 1 hypothesis\\ 
  - 2 = multiple occurrences in 1 hypothesis\\ 
  - 3 = single occurrence in 2 hypotheses\\ 
  - 4 = multiple occurrences in 2 hypotheses\\ 
  - 5 = present in all 3+ hypotheses\\ 

\textbf{\#\#\# RF-09 — Failure to update belief (non-monotonic updating error)} \\

\textbf{Scope}: full-history\\ 
\textbf{Definition}: Model does not revise or retract a previous claim after later evidence or Tool Message contradicts it. This is specifically a failure to update in light of new evidence (distinct from anchoring, which is failure to explore alternatives).\\ 
\textbf{Example}: Claims Redis eviction; later Tool Message shows normal Redis; final answer still claims eviction. Speculates webservice2 is hosted on host4 and therefore failures on host4 affect webservice2; later Tool Message shows webservice is hosted on host2; final answer still includes basis of webservice2 being hosted on host4.\\ 
\textbf{Signals}: Later Tool Messages/alerts contradict earlier claims, and claims persist.\\ 
\textbf{Annotation rule}: Extract the original claim and the contradicting evidence; if model fails to revise → mark RF-09. Severity high if final answer relies on unchanged, contradicted claim.\\ 
\textbf{Severity}: 1-5\\ 
  - 1 = single untrue claim impacts 1 hypothesis\\ 
  - 2 = multiple impact 1 hypothesis\\ 
  - 3 = single impacts 2 hypotheses\\ 
  - 4 = multiple impact 2 hypotheses\\ 
  - 5 = untrue claim(s) impact all 3 hypotheses\\ 

\textbf{\#\#\# RF-10 — Simulation / role confusion (pretend tool output used as factual evidence)} \\

\textbf{Scope}: full-history\\ 
\textbf{Definition}: Model explicitly states it cannot call tools and "assumes" or "simulates" tool outputs, then treats simulated outputs as factual in final conclusions. Consider a Tool Message "real" if there is an explicit tool header like "==== Tool Message ====" (allowing variable `=' counts).\\ 
\textbf{Example}: "I cannot call the tools; I will assume the log shows ERROR: connection refused"; final answer treats the assumed log as observed.\\ 
\textbf{Signals}: Phrases like: "I cannot call", "I can't call", "I'll assume", "I will pretend", "simulate the response", "assume the tool returns", followed by conclusive claims. No Tool Message corresponding to the assumed call.\\ 
\textbf{Annotation rule}: Consider all text prior to the final answer and search for signal phrases and Tool Messages. If simulated outputs are used as factual evidence without a Tool Message, mark RF-10. If simulated exploration was not used as final evidence or Tool Messages were later present, mark RF-10 but lower severity.\\ 
\textbf{Severity}: 1-5\\ 
  - 1 = simulated output noted, but not used for final claims\\ 
  - 2 = simulated output used for a secondary claim only\\ 
  - 3 = simulated output(s) used as evidence for 1 hypothesis\\ 
  - 4 = simulated output(s) across 2 hypotheses\\ 
  - 5 = simulated output(s) are the core evidence for 3+ hypotheses\\ 

\textbf{\#\#\# RF-11 — Excessive speculative / rambling reasoning (ungrounded token waste)} \\

\textbf{Scope}: full-history\\ 
\textbf{Definition}: Considerable portion of the chat is spent being confused or speculating about the system architecture, knowledge graph, meaning semantics of the alerts, available tools, and deciding what steps to take instead of performing tool calls to confirm/refute KG characteristics or check evidence; especially when tools were available but unused.\\ 
\textbf{Example}: KG-schema theorizing while no Tool Messages are present to confirm/refute KG characteristics.\\ 
\textbf{Signals}: High density of hedging or rambling language (including "wait"); paragraphs without alert/metric/log/trace citations; round-about or circular thoughts; none-to-little Tool Message usage.\\ 
\textbf{Annotation rule}: Consider all text prior to the final answer and search for hedging or speculative language. If high and blocked/replaced necessary data checks OR prevented a conclusive unstructured final answer, mark RF-11.\\ 
\textbf{Severity}: 3-5\\ 
  - 3 = heavy speculation/rambling, blocked some important checks\\ 
  - 4 = very heavy speculation/rambling, significantly blocked analysis and tests\\ 
  - 5 = entire session dominated by speculation/rambling, no meaningful evidence work, final answer driven by speculation\\ 

\textbf{\#\#\# RF-12 — Repetition / failure to resume (looping across turns)} \\

\textbf{Scope}: full-history\\ 
\textbf{Definition}: Model repeats the same planning, intro text, or deliberation across consecutive replies and fails to resume earlier progress,typically after truncation.\\ 
\textbf{Example}: Consecutive replies (marked by "=== AI Message ===") begin with similar "First I will check…" paragraphs and add little new content. Consecutive replies contain similar deliberation about the semantics of a "down"/"up" metric alert.\\ 
\textbf{Signals}: High n-gram overlap across AI messages; or repeated planning text.\\ 
\textbf{Annotation rule}: If repetition caused stalled progress or omitted checks, mark RF-12.\\ 
\textbf{Severity}: 3-5\\ 
  - 3 = repetition across multiple turns omitted some checks\\ 
  - 4 = repetition stalled significant parts of the analysis\\ 
  - 5 = repetition prevented completion and changed final answer\\ 

\textbf{\#\#\# RF-13 — Anchoring / premature commitment (insufficient hypothesis exploration)} \\

\textbf{Scope}: cross-cutting (search behaviour)\\ 
\textbf{Definition}: Model fixates early on a single hypothesis and fails to enumerate OR to explore other plausible alternatives/hypotheses (e.g., component or type of fault).\\ 
\textbf{Example}: Model immediately focuses on "high memory usage" as the cause and never lists or considers other plausible causes (e.g., network, disk) despite relevant alerts. Model claims it should explore host relationships for loginservice2, webservice1, and dbservice2; calls the tool for loginservice2; does not follow-through for webservice1 and dbservice2.\\ 
\textbf{Signals}: $<2$ reasonable alternatives (w.r.t. component or fault type) listed across chat; no follow-through on planned exploration without good rationale.\\ 
\textbf{Annotation rule}: If the model provides fewer than 2 reasonable alternative hypotheses and goes straight to a definitive cause, mark RF-13.Only consider RF-13 if there was an opportunity to explore (i.e., through tools, using the reasoning fields, think tags \texttt{<think\></think\>}, etc.).\\ 
\textbf{Severity}: 3-5\\ 
  - 3 = some diversity in planned exploration and some follow-through\\ 
  - 4 = some diversity in planned exploration but no follow-through \\ 
  - 5 = anchoring dominated the analysis and impacted all hypotheses \\ 

\textbf{\#\#\# RF-14 — Invalid logical inference patterns (formal fallacies)} \\

\textbf{Scope}: cross-cutting (invalid inference)\\ 
\textbf{Definition}: Model applies invalid inference patterns in deriving diagnostic claims. Look for: affirming the consequent, denying the antecedent, post hoc, composition/division, ecological fallacy, hasty generalization.\\ 
\textbf{Example}: Model sees one trace with a 500 for endpoint /login on a single instance and concludes "the whole service is down" (hasty generalization).\\ 
\textbf{Signals}: Clear leap from limited premises to broad/systemic conclusion without intermediate mechanism or checks.\\ 
\textbf{Annotation rule}: Identify fallacy, quote the premise(s) and conclusion, mark RF-14.\\ 
\textbf{Severity}: 1-5\\ 
  - 1 = single minor fallacy with negligible impact\\ 
  - 2 = multiple minor fallacies with negligible impact\\ 
  - 3 = fallacy(s) used to support 1 hypothesis\\ 
  - 4 = fallacy(s) used to support 2 hypothesis\\ 
  - 5 = fallacies pervasive across all 3+ hypotheses\\ 

\textbf{\#\#\# RF-15 — Internal contradiction (explicit inconsistency)} \\

\textbf{Scope}: cross-cutting\\ 
\textbf{Definition}: Model makes mutually incompatible statements in the chat history. This is different from RF-09: RF-15 is explicit contradiction rather than failure to revise.\\ 
\textbf{Example}: "No trace 500 errors", then "multiple 500 trace errors".\\ 
\textbf{Signals}: Pairwise contradictory sentences when compared; contradiction can be related (but not limited) to a metric, timestamp, evidence existence, component relationships, etc.\\ 
\textbf{Annotation rule}: Quote contradictions, mark RF-15.\\ 
\textbf{Severity}:\\ 
  - 1 = single contradiction with negligible impact\\ 
  - 2 = multiple contradictions with negligible impact\\ 
  - 3 = contradiction(s) impact 1 hypothesis\\ 
  - 4 = contradiction(s) impact 2 hypotheses\\ 
  - 5 = contradiction(s) impact 3+ hypotheses\\ 

\textbf{\#\#\# RF-16 — Arithmetic / aggregation mistake} \\

\textbf{Scope}: cross-cutting\\ 
\textbf{Definition}: Numeric miscalculations or wrong aggregations that change interpretation.\\ 
\textbf{Example}: Reports "error rate increased 200\%" when correct is 20\%.\\ 
\textbf{Signals}: Numeric expressions in the text; automatic recomputation disagrees with reported number.\\ 
\textbf{Annotation rule}: Recompute numeric or aggregation claims; if inconsistent and materially affects conclusions, mark RF-16 + include corrected value. (Severity generally low unless numeric error changed final diagnosis.)\\ 
\textbf{Severity}: 1-5\\ 
  - 1 = single small numeric mismatch with negligible impact\\ 
  - 2 = multiple minor numeric mismatches with negligible impact\\ 
  - 3 = numeric/aggregation error(s) that impacts 1 hypothesis\\ 
  - 4 = numeric/aggregation error(s) that impact 2 hypotheses\\ 
  - 5 = numeric/aggregation error(s) that impact 3+ hypothesis\\ 

\textbf{\#\# TASK } \\
Your task is to label the given reasoning by an LLM agent below according to the failure taxonomy.\\ 

\textbf{\#\#\# ANNOTATION GUIDE -- practical rules} \\ 
1. Multilabel: Assign all RFs that apply.\\ 
2. RF-00 precedence: If the structured final response (JSON) is None or nan, do NOT mark RF-00. Otherwise, compare the unstructured final answer with the structured final response; if divergent, mark RF-00 and use the unstructured final answer as the basis for the rest of the annotation.\\ 
3. Tool message rule: If the chat contains a real Tool Message corresponding to the claimed tool call → treat associated evidence as real. If the model claimed a call but no Tool Message exists and the model used the assumed output as fact → consider RF-10. Consider a Tool Message "real" if there is an explicit tool header like "==== Tool Message ====" (allowing variable `=' counts).\\ 
4. Document evidence: For every RF, paste the triggering model sentence(s) and the matched alert(s) or NO MATCH FOUND.\\ 
5. Severity per RF (1-5): Give severity for each RF assigned.\\ 
6. Ground-truth is context only: Do not mark RFs solely because the model disagrees with ground truth; only flag when model claims/presents evidence incorrectly relative to accessible context.\\ 

\textbf{\#\#\# ANNOTATION WORKFLOW (step-by-step)} \\ 

\textbf{Step 1 — Global gate} \\ 
1. Compare final structured output response vs final unstructured answer → mark RF-00 if divergent; record severity.\\ 

\textbf{Step 2 — Per-hypothesis loop (for each hypothesis in the final answer, process top-1 first)} \\
2. RF-01: quick-scan up to 3 times for EACH ; if no match → RF-01 for that hypothesis → stop per-hypothesis checks for *this* hypothesis.\\ 
3. RF-02: if metric alerts used, verify semantics; if inverted/misread → RF-02.\\ 
4. RF-03: examine provenance; if model blames observer despite propagation language → RF-03.\\ 
5. RF-04: compare timestamps; if cause occurs after effect → RF-04.\\ 
6. RF-05: evaluate causal mechanism vs KG/alerts; if mechanism relies on non-existent KG edges or is otherwise unsupported/not plausible → RF-05.\\ 
7. RF-06: if model claims instance-level root cause location, verify unique per-instance multi-modal evidence; if absent → RF-06.\\ 
8. RF-07: assess whether chosen evidence selection contradicts simple triage heuristics and whether selection changed conclusions; if so → RF-07 (severity scaled by impact).\\ 
9. RF-08: apply Sufficiency Checklist (Temporal, Frequency, Mechanism, Discriminability, Provenance, Granularity); if required items fail for claim type → RF-08.\\ 

\textbf{Step 3 — Full-history checks (scan all "AI Message" and "Tool Message" instances prior to the Final Answer/Response)} \\ 
10. MANDATORY: perform a full history scan and search for evidence for RF-09-RF-12. Produce a compact "Full History Summary" of the LLM agent's behaviour. Provide exact quote(s) (verbatim) to support RF-09-RF-12.\\ 
11. RF-09: scan chronological history for later Tool Messages/alerts that contradict earlier claims; if no revision → RF-09.\\ 
12. RF-10: detect simulated tool outputs used as facts without Tool Message → RF-10.\\ 
13. RF-11: measure speculative text percentage; if $>30$\% and blocked checks → RF-11.\\ 
14. RF-12: detect repeated planning text that stalled progress → RF-12.\\ 

\textbf{Step 4 — Cross-cutting checks (scan the entire chat history)} \\ 
15. MANDATORY: perform an entire chat history scan and search for evidence for RF-13-RF-16.\\ 
16. RF-13: did the agent ever enumerate $\geq$2 plausible alternatives? If not → RF-13.\\ 
17. RF-14: detect formal fallacies anywhere → RF-14.\\ 
18. RF-15: find explicit contradictions elsewhere → RF-15.\\ 
19. RF-16: recompute numeric claims or aggregations across chat; if mismatches materially affect reasoning → RF-16.\\ 

\textbf{Step 5 — Finalize }\\ 
20. Assign all applicable RFs and record per-RF severity (1-5). For the per-hypothesis RFs, make sure the severity accurately reflects the number of occurrences across hypotheses.\\ 
21. Out of the RFs identified, list the RFs that directly affected/impacted the \#1 hypothesis.\\ 
22. Output a json object (using (\texttt{```json}) and (\texttt{```}) as delimiters) with the Failures Identified Output Schema.\\ 

\textbf{\#\#\# FAILURES IDENTIFIED OUTPUT SCHEMA} \\

\noindent\begin{minipage}{\linewidth}
\begin{verbatim}
'''json
{
  "failures_identified": [
    {
      "type": "The RF identifier, e.g. 'RF-01'",
         "model_claim": "Model claim or behaviour in the chat history the supports
                     the RF",
       "rationale": "A description of the issue and a justification/rationale 
                   that the RF applies",
      "severity": "Severity of the RF."
    },
    ...
  ],
  "affected_top_hypothesis": "List of RFs that directly affected the 
                             #1 hypothesis, e.g., ['RF-01', 'RF-13']"
}
'''
\end{verbatim}
\end{minipage}

Below is the original root cause analysis task and the associated "final response" (the LLM's structured output), all enclosed in \texttt{<begin chat history>} and \texttt{<end chat history>}.\\ 
The ground-truth root cause for this scenario was: \{\textit{root cause location}\} (location) and \{\textit{root cause type}\} (type).\\ 

Work through the annotation workflow step by step.\\ 

\texttt{<begin chat history>}\\ 

\{\textit{chat history}\}\\ 

Final response (structured output):\\ 
\{\textit{final answer}\}\\ 

\texttt{<end chat history>}
\end{tcolorbox}
\captionof{figure}{LLM-as-a-Judge prompt template with the reasoning failure taxonomy and step-by-step annotation workflow.}
\label{fig:judge-prompt}

\end{document}